\documentclass[12pt, epsfig]{article}

\usepackage{amsfonts}
\usepackage{amssymb}
\usepackage{graphics}
\usepackage{epsfig}
\usepackage{multirow}

\newcommand{\beq}{\begin{equation}}
\newcommand{\eeq}{\end{equation}}
\newcommand{\beqn}{\begin{eqnarray}}
\newcommand{\eeqn}{\end{eqnarray}}
\newcommand{\bea}{\begin{eqnarray}}
\newcommand{\eea}{\end{eqnarray}}
\newcommand{\beas}{\begin{eqnarray*}}
\newcommand{\eeas}{\end{eqnarray*}}

\newcommand{\bquo}{\begin{quote}}
\newcommand{\enqu}{\end{quote}}

\def\qt{\widetilde{q}}
\def\vp{\varphi}
\def\Tr{ \hbox{\rm Tr}}

\def\stroke{\vrule height8pt width0.4pt depth-0.1pt}
\def\topfleck{\vrule height8pt width0.5pt depth-5.9pt}
\def\botfleck{\vrule height2pt width0.5pt depth0.1pt}
\def\Zmath{\vcenter{\hbox{\numbers\rlap{\rlap{Z}\kern 0.8pt\topfleck}\kern
2.2pt
                   \rlap Z\kern 6pt\botfleck\kern 1pt}}}
\def\Qmath{\vcenter{\hbox{\upright\rlap{\rlap{Q}\kern
                   3.8pt\stroke}\phantom{Q}}}}
\def\Nmath{\vcenter{\hbox{\upright\rlap{I}\kern 1.7pt N}}}
\def\Cmath{\vcenter{\hbox{\upright\rlap{\rlap{C}\kern
                   3.8pt\stroke}\phantom{C}}}}
\def\Rmath{\vcenter{\hbox{\upright\rlap{I}\kern 1.7pt R}}}
\def\Z{\ifmmode\Zmath\else$\Zmath$\fi}
\def\Q{\ifmmode\Qmath\else$\Qmath$\fi}
\def\N{\ifmmode\Nmath\else$\Nmath$\fi}
\def\C{\ifmmode\Cmath\else$\Cmath$\fi}
\def\R{\ifmmode\Rmath\else$\Rmath$\fi}

\def\Tr{{\rm Tr}}

\def\2{{1\over 2}}

\def\ca{\cos{\alpha\,}}
\def\sa{\sin{\alpha\,}}
\def\4N{${\cal N}=4$}
\def\N{{\cal N}}

\def\beq{\begin{equation}}
\def\eeq{\end{equation}}
\def\ba{\beq\new\begin{array}{c}}
\def\ea{\end{array}\eeq}

\def\U{{\rm U}}

\newcommand{\gs}{g^{2}}

\newcommand{\ve}{\varepsilon}

\newcommand{\pz}{\partial_{z}}

\begin{document}

\begin{titlepage}

\begin{flushright}
FTPI-MINN-08/24; UMN-TH-2703/08\\

\end{flushright}

\vspace{1mm}

\begin{center}
{\large  {\bf Confinement and Localization on Domain Walls }}
\end{center}

\vspace{1mm}

\begin{center}
{\large  R. {\sc Auzzi}$^{(1)}$}, {\large S. {\sc Bolognesi}$^{(2)}$}, 
{\large M. {\sc Shifman}
$^{(2,3)}$}  and \\ {\large  A. {\sc Yung}$^{(2,4)}$} \vskip 0.20cm
\end{center}
\begin{center}
$^{(1)}${\it \footnotesize
Department of Physics, Swansea University, \\ Singleton Park, Swansea SA2 8PP, U.K.}
 \\ \vskip 0.20cm  $^{(2)}${\it \footnotesize
William I. Fine Theoretical Physics Institute, University of Minnesota, \\
116 Church St. S.E., Minneapolis, MN 55455, USA}\\ \vskip 0.20cm  
$^{(3)}$ {\it \footnotesize
Laboratoire de Physique Th\'eorique\footnote{Unit\'e
Mixte de Recherche du CNRS,  (UMR 8627).}
Universit\'e de Paris-Sud XI\\
B\^atiment 210, 
F-91405 Orsay C\'edex, FRANCE
}
\\ \vskip 0.20cm
$^{(4)}$
{\it \footnotesize Petersburg Nuclear Physics Institute, Gatchina, St. Petersburg
188300, RUSSIA}\\
{\it \footnotesize and\\
 Institute of Theoretical and Experimental Physics, Moscow
117259, RUSSIA}
\end {center}
\vspace{1mm}

\begin{abstract}

We continue the studies of localization of the
U(1) gauge fields on domain walls.
Depending on dynamics of the bulk theory 
the gauge field localized on the domain wall can be either in the Coulomb phase or
squeezed into flux tubes implying (Abelian) confinement of probe charges
on the wall along the wall surface.
First,  we consider a simple toy model with one flavor 
in the bulk at weak coupling (a minimal model) 
realizing the latter scenario.  We then suggest a model presenting an extension of the Seiberg--Witten theory which is at strong coupling, but all theoretical constructions are under full
control if we base our analysis on a dual effective action.
Finally, we compare our findings with the  wall
in a ``nonminimal" theory with two distinct quark flavors that had been studied previously. 
In this case the U(1) gauge field trapped on the wall is exactly massless because
it is the Goldstone boson
of a U(1) symmetry in the bulk
spontaneously broken on the wall. 
The theory on the wall is in the Coulomb phase.
 We explain why  the mechanism of confinement 
 discussed in the first part of the paper
 does not work in this case, and  strings  are not formed on the walls. 

\end{abstract}

\end{titlepage}

\section{Introduction}
\label{intro}

Localization of gauge fields on domain walls which are supported by some
four-dimensional gauge theories is discussed in the literature for a long time
\cite{Dvali:1996xe, Acharya:2001dz,Dubovsky:2001pe,Shifman:2002jm, Auzzi:2006ju,
Dvali:2007nm}. Elementary domain walls localize U(1) fields. As was explained by Polyakov
\cite{Polyakov:1976fu}, in 2+1 dimensions the U(1) gauge field is dual
to a phase field $\sigma$ living on $S_1$. The U(1) gauge theory in 2+1 dimensions
can exist in distinct regimes: (i) Coulomb, with the long-range interaction $\ln r$
due to the exchange of the gauge field; (ii) the gauge field is Higgsed, electric charges are screened,
interaction  due to the exchange of the gauge field falls off exponentially; (iii)
the gauge field acquires a mass through the Chern--Simons term,
gauge symmetry is unbroken; and (iv) the dual photon field $\sigma$ gets a mass term.
This latter regime is quite peculiar. It might seem that the mass term of
the $\sigma$ field implies short-range interactions. In fact, it is the opposite!
Electric charges (seen as the $\sigma$ field vortices in the dual language)
are connected by a flux line which plays the role of a confining string.
Interaction between the electric charges grows linearly with the distance $r$.
In terms of $\sigma$ the string is a domain line very similar to the axion domain walls
in 3+1 dimensions.
The domain line endpoints are the $\sigma$ field vortices. (For some reviews see
Refs.~\cite{Shifman:2007ce,Tong:2005un,Eto:2006pg}.)

The domain lines of the $\sigma$ field are the essence of the
Polyakov confinement \cite{Polyakov:1976fu}. Polyakov's model is 2+1 dimensional compact electrodynamics. It represents the low-energy limit of SU(2) Yang--Mills theory
with one adjoint Higgs field which develops a vacuum expectation value (VEV)
breaking SU(2) down to U(1). The mass term for the dual photon is generated by
SU(2) three-dimensional instantons (in 2+1 dimensions, technically, they are identical to 
't~Hooft--Polyakov monopoles \cite{tHP}). When the U(1) gauge field
is dynamically localized on a wall occurring in 3+1 dimensional theory,
which of the four regimes listed above is in fact implemented depends on details of the bulk
theory.

The first example of a U(1) gauge field localized on a wall, in a 
fully controllable theoretical setting, was given in \cite{Shifman:2002jm}.
In this example a global U(1) symmetry of the bulk theory, spontaneously broken
on the wall, guarantees masslessness of the 2+1 dimensional
photon.\footnote{The dual photon $\sigma$ is the Goldstone field.} The U(1) theory on the wall is in the Coulomb regime. When the global U(1) symmetry is explicitly weakly broken in the bulk,
the  $\sigma$ field becomes quasi-Goldstone, a $\sigma$ mass term is generated
implying confinement of the electric charges on the wall \cite{Auzzi:2006ju}.

In a recent paper~\cite{Dvali:2007nm} a mechanism (developing a concept put forward in
\cite{Dvali:1996xe})
has been suggested that  leads to confinement on domain walls. 
Unlike the models discussed in \cite{Shifman:2002jm, Auzzi:2006ju},
consideration of Ref.~\cite{Dvali:2007nm} was carried out in nonsupersymmetric setting, 
although the mechanism {\it per se} is general and can be implemented in 
a wide class of bulk theories, both supersymmetric or nonsupersymmetric.
The only requirements to these theories are: they should support both
domain walls and Abrikosov--Nielsen--Olesen (ANO) flux tubes \cite{ANO} and be minimal 
(in which sense minimal will be explained later). Far away from the wall the charged field
condensate responsible for the ANO flux tubes is ``large" 
and squeezes the flux tube from all directions in the perpendicular plane.
Now, if we place such tube inside the wall, where the above condensate vanishes with an exponential accuracy, in the first approximation the confining regime gives
place to the  Coulomb regime on the wall.
The flux tube is still squeezed inside the wall in the direction perpendicular to the wall; however, it swells in the directions parallel to the wall.

In the next approximation one should take
into account the fact that there is a residual charged field 
condensate inside the wall. Although it is exponentially small, it still
limits the swelling of the flux tube placed inside the wall in the directions parallel to the wall.
The thickness of the flux tube in these directions is exponentially large, but finite. 
If we go to still larger distances along the wall, (magnetic)
charges attached to the endpoints of such a tube
experience linear confinement.

\begin{figure}[h!tb]
\epsfxsize=10cm \centerline{\epsfbox{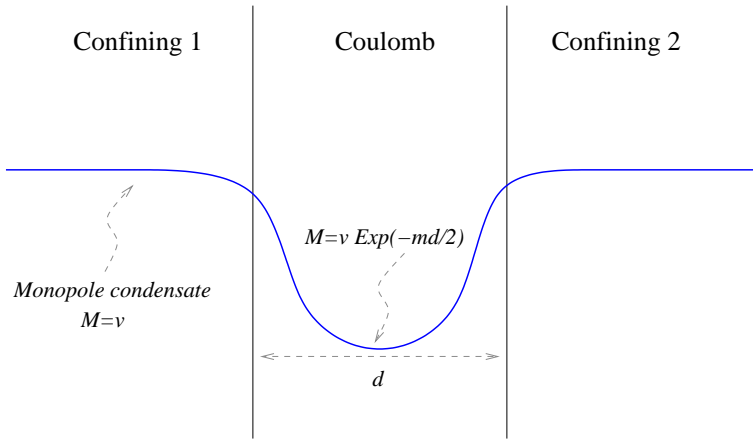}}
\caption{{\protect\small The condensate profile for the wall of the type 
Confining 1--Coulomb--Confining 2.}} 
\label{monopoleprofile}
\end{figure}

The above description is phrased in terms of the charged field condensate
and magnetic flux tubes. Needless to say, in actuality we keep in mind a dual picture,
presented in Fig.~\ref{monopoleprofile}: the monopole condensation
leading to electric flux tubes. In what follows the dualization will be tacitly assumed.
Thus, when we speak of matter fields that condense, we will keep in mind that
these local fields present an effective description of monopoles, 
much in the same way as in the Seiberg--Witten construction \cite{Seiberg:1994rs}.

The suggestion put forward in \cite{Dvali:2007nm} is inspirational.
At the same time, operational mode of this mechanism remained unclear,
as well as its relation to other regimes
implementable in the models with the U(1) gauge field localized on the walls.
Moreover, particular models considered in \cite{Dvali:2007nm} 
suffer from the wall-antiwall instability.
With these instabilities, working out quantitative details does not seem possible.

The
purpose of the present paper is to 
address these issues. We focus on investigation of how this mechanism actually works,
and what deformations or modifications lead to deconfinement.
We suggest two stable model examples: one at weak coupling and another using the 
Seiberg--Witten solution \cite{Seiberg:1994rs} at strong coupling. These models are demonstrated to
be working examples of confinement on the domain wall. {\em En route}, we will also clarify some aspect regarding localization of the gauge fields on the domain walls. 

To ensure stability of the model it is necessary to require that
two vacua in which the matter fields condense are two {\em distinct} vacua. Let us call them
Confining 1 and Confining 2 (Fig.~\ref{monopoleprofile}).\footnote{In the dual language
Confining 1 and Confining 2 read Higgs 1 and Higgs 2.}
 If the monopole mass 
in the Coulomb phase is $m$, the condensate in the center of the wall
is roughly $ve^{-md/2}$ where $d$ is the wall thickness.
If $md$ is large enough, the condensate inside the wall almost vanishes, 
and the gauge
theory exists inside the wall in the (almost) Coulomb regime.
Deviations of this almost Coulomb regime from the perfect Coulomb regime
determine the thickness of the flux tube in the directions parallel to the wall.

The exponentially small deviation from the perfect Coulomb regime inside the wall
has a clear-cut interpretation in terms of the field $\sigma$.
Instead of being the modulus field, as in \cite{Shifman:2002jm}, it becomes a quasimodulus.
We work out a method which allows one to calculate its
mass directly from the bulk theory. Moreover, introduction of the second matter field,
as in \cite{Shifman:2002jm}, can restore the exact modulus status of $\sigma$.

The paper is organized as follows. 
In Sect. 2 we discuss some conceptual aspects.
Section \ref{toymodel} introduces a ``minimal"
model,  the simplest example which realizes the confinement mechanism discussed by Dvali {\em et al.} Special attention is given 
here to discussion of the localization mechanism for the gauge field
and  formation of the flux tubes in the wall. 
This example is quasiclassical (the model is weakly coupled) and is phrased in dual terms.
The quark field condenses while the monopoles are confined by the Abrikosov--Nielsen--Olesen string. Section \ref{seibergwitten} provides an example at strong coupling in which condensation that occurs is that
of the monopole field. This example is an extension of  $\N=2$ super-Yang--Mills
theory (slightly broken to $\N=1$) considered by Seiberg and Witten.
The Seiberg--Witten solution is an essential ingredient
which allows us to treat the theory at strong coupling. In  Sect.~\ref{2flavors} we consider
a ``nonminimal" theory with two flavors and explain why, in contradistinction with
the minimal model of Sect.~2, the gauge field on the wall remain massless despite the presence of a residual condensate in the middle of the wall. 
The theory on the wall is in the Coulomb regime. 
The masslessness is backed up by the Goldstone theorem:
in the bulk we have an exact global U(1) symmetry which is spontaneously
Section \ref{conclusion} 
summarizes our findings.

\section{Conceptual aspects}
\label{CA}

The basic idea of the gauge field localization suggested in \cite{Dvali:1996xe}
assumes that the bulk four-dimensional theory is in the confining regime
while inside the 1+2-dimensional wall we have ``less confinement" (or no confinement at all).
Then the chromoelectric flux coming from the bulk through a tube spreads out inside the wall.
The flux tube-wall junction plays the role of the color source inside the wall.
In the dual formulation the bulk theory is Higgsed while inside the wall  it is ``less
Higgsed" (not Higgsed at all in the case of U(1)). 

The technical implementation of this idea is not quite straightforward.
Indeed, say, in the U(1) case which has just been mentioned
the magnetic charges are confined in the bulk. Thus,  the magnetic flux 
from a distant magnetic monopole is squeezed into a tube, and when this tube hits the wall, it
spreads out inside the wall. To describe this phenomenon in terms of the standard 
1+2-dimensional U(1) gauge theory  {\em on} the wall surface we have to use a 
duality transformation which converts the magnetic field inside the wall into
a dual electric field of the effective theory on the wall surface.
The flux tube-wall junction acts as an {\em electric} charge source in 
1+2-dimensional electrodynamics on the wall. This duality transformation is a crucial element
of the construction. The relation between the gauge potential in the bulk and that
in the effective low-energy theory on the wall surface is nonlocal.

In 1+2 dimensions the gauge field $A_\mu$ is dual to a phase field $\sigma$
\cite{Polyakov:1976fu}. If a U(1) symmetry is an exact symmetry of the bulk theory, spontaneously broken on the wall, occurrence of the Goldstone field $\sigma$ localized on the wall
is inevitable. Dualizing the above Goldstone field we get massless electrodynamics on the wall.
The electric flux from a charge source in the effective
theory on the wall surface is spread according to the Coulomb law.
If, on the other hand, a global U(1) symmetry is only an approximate symmetry of the 
bulk theory \cite{Auzzi:2006ju}, or even just  an approximate symmetry of
the domain wall solution \cite{Dvali:1996xe}, we should expect a {\em pseudo}-Goldstone mode localized on the wall, with a small mass term. With this pseudo-Goldstone mode we get electrodynamics with confinement
on the wall, 
\`a la Polyakov. The electric flux of the world-volume theory is,
in its turn, squeezed on the wall, forming a band of thickness inversely proportional
to the {pseudo}-Goldstone mass. This thickness is exponentially larger than that
of the bulk magnetic flux tubes.

These two scenarios are realized in two-flavor and one-flavor models, respectively.

\section{The Simplest Example at Weak Coupling}
\label{toymodel}

\subsection{Theoretical Setting}

To introduce the reader to the subject we will start with a
toy model that contains all relevant features of the physical phenomenon
we want to describe. Consider a U$(1)$ gauge theory with a
charged scalar field $Q$. Our task is to
study a  domain wall interpolating between Higgs--Coulomb--Higgs vacua. 
The model is nonsupersymmetric. The simplest choice of the potential seems to be as follows:
\beq
|Q|^2(|Q|^2 -v^2)^2\,,
\label{pote}
\eeq
as suggested in \cite{Dvali:2007nm}. However, there is a problem
with (\ref{pote}), namely  
the  Higgs--Coulomb--Higgs interpolation
is a wall-antiwall configuration in this model, which is unstable.
Of course, the instability can be made exponentially small, but so are the effects
we try to trace.

\begin{figure}[h!tb]
\epsfxsize=7cm \centerline{\epsfbox{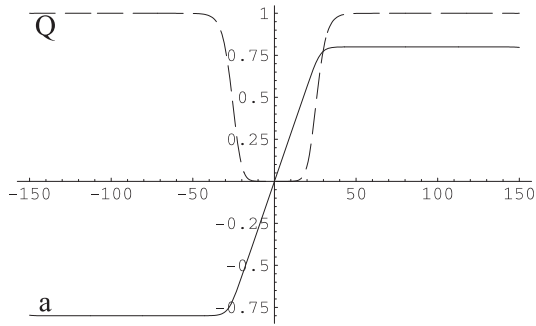}}
\caption{{\protect\small Domain wall in the  toy model (\ref{lagr}), (\ref{potep}).
The profile of the $a(z)$ field is presented by the
solid line while that of $Q(z)$ by dashed. The horizontal axis presents the direction $z$ perpendicular to the wall.}} 
\label{pl1}
\end{figure}

To create a stable configuration we need at least an extra real
neutral field. Consider a system in the Higgs phase, with the
U$(1)$ gauge group, a scalar field $Q$ with charge $+1$, and an
uncharged scalar $a$, with the following Lagrangian:
\beq
L=-\frac{1}{4e^2} (F^{\mu \nu})^2+ \frac{1}{2 e^2}
(\partial_\mu a)^2+ |\nabla_\mu Q|^2 - V \,,
\label{lagr}
\eeq
with the potential
\beq 
V=\frac{(a-m)^2 (a+m)^2 |Q|^2}{m^2} +
 \frac{e^2}{2} (v^2-|Q|^2)^2\,.
 \label{potep} 
 \eeq
This model is non-renormalizable, but we can still
consider it as an effective theory in the infrared.\footnote{This
example can be embedded in supersymmetric QED (SQED) with the Fayet--Iliopolous
$D$-term, the superpotential  $$ W= Q
\tilde{Q} \,\,\frac{(a-m)(a+m)}{m}\, ,$$
and in a vacuum with $\tilde{Q}=0$. } 
It is obvious then that we have two distinct vacua
$a= \pm m$, $|Q|=v$.
If we keep $m\gg e v$,
there is a large intermediate region inside the domain wall where the
VEV of $Q$ is almost zero (see Fig.~\ref{pl1}). Even if the
theory has no Coulomb vacuum, if we look at the domain wall profile
we immediately see that this inside region of the wall is almost in the Coulomb
phase. In the limit $m \rightarrow \infty$ the wall becomes
infinitely thick, and the Higgs VEV in the wall center  tends
to zero.

To see this we write down the equations of motion for our model.
Since we are looking for the domain wall solution we 
assume that all fields are static and depend only on a single spatial coordinate $z$
(the wall is perpendicular to the $z$-axis), and drop the gauge field. We have
\beqn
&&
\pz^2 a = 4a\,\frac{a^2-m^2}{m^2}\,Q^2\,,
\nonumber\\[3mm]
&&
\pz^2 Q - Q\,\frac{(a^2-m^2)^2}{m^2} - e^2(|Q|^2-v^2)\,Q\,=0\,.
\label{dwalleqs}
\eeqn
In the leading approximation the quark field $Q$ vanishes inside the wall. Then the first equation above gives for 
the neutral scalar $a$
\beq
a\approx 2m\,\frac{z-z_0}{d},
\label{asol}
\eeq
where $z_0$ is the center of the wall, while $d$ is its
thickness, to be determined below. In Eq.~(\ref{asol}) we take into
account the $a$-field boundary conditions, $a\to\pm m$ at
$z\to \pm\infty$.
Substituting (\ref{asol}) in the second equation in
 (\ref{dwalleqs}) we can write the following equation for the profile field $Q$,
neglecting the non-linear terms:
\beq 
\pz^2 Q  =  Q\,
\left\{\frac{16m^2}{d^4}\left[(z-z_0)^2 -\frac{d^2}{4}\right]^2 -e^2v^2
\right\} 
\, . 
\label{lineariz}
\eeq

Below we will show that $d\sim m/e^2v^2$. Taking this into account, 
inside the wall, near its left edge,
\beq
1/ev\ll z-z_0+\frac{d}{2}\ll d\,,
\label{leftedge}
\eeq
we can simplify the above  equation by dropping the second term in the 
curly brackets and replacing $z-z_0 -d/2$ by $-d$. Then we get
\beq
\pz^2 Q  =  Q\,\frac{16m^2}{d^2}\left(z-z_0+\frac{d}{2}\right)^2 .
\eeq
The solution of this equation obviously has the form
\beq
Q\approx v\,e^{-\frac{2m}{d}(z-z_0+\frac{d}{2})^2},
\label{Qlprofile}
\eeq
where we use Eq.~(\ref{leftedge}).

Much in the same way, near the right edge of the wall at 
\beq
1/ev \ll -\left( z-z_0-\frac{d}{2}\right) \ll d
\eeq
we derive from (\ref{dwalleqs})
\beq
Q\approx v\,e^{-\frac{2m}{d}(z-z_0-\frac{d}{2})^2}\,.
\label{Qrprofile}
\eeq
Note that these quark profiles are similar to those obtained in 
\cite{Shifman:2002jm} in supersymmetric QED with two quark
flavors, see Sect.~4.

Let us estimate $d$ in the limit $m\gg e v$.
In the inside region the VEV of $Q$ almost vanishes while $a$
is linear in $z$. In order to estimate the thickness of this region
(i.e. the wall thickness), let us first
estimate the wall tension as a function of $d$ and then minimize it with respect to $d$,
\beq 
T_{\rm wall} \sim \frac{e^2 v^4}{2} d + \frac{(2m)^2}{d e^2}\,. 
\label{tensw}
\eeq
The assumptions for this estimation are that for $m\gg e v$
the dominating contributions to the energy come from the 
potential and  kinetic terms of  the  $a$ field.
The minimum is achieved at
\beq 
d= \frac{2 \sqrt{2} m}{e^2 v^2} \, ,
\label{ma}
\eeq
where $ev$ is the photon mass in the bulk.
The tension of the wall is of  the order of 
\beq 
T_{\rm wall} \sim 2 \sqrt{2} \, m\,v^2. 
\label{tw}
\eeq

In what follows, we shall be interested in trapping gauge fields, inside this domain wall.
Localization of gauge fields on lower-dimensional topological objects 
is, generally speaking, a
nontrivial task. Massless scalars can be localized as Goldstone bosons of
continuous symmetries spontaneously broken on the given topological defects.
Massless fermions can be localized via Jackiw--Rebbi's and other index 
theorems~\cite{Jackiw:1975fn} (see \cite{Indexth,Gorsky:2007ip}).
We shall discuss this issue in more  detail in Section \ref{dmvqm}.

As discussed in Ref.~\cite{Shifman:2002jm}, we can consider the
following gauge invariant order parameter:
\beq 
\label{order}
e^{i \sigma} = v^{-2} \bar{Q}(-\infty) e^{i \int A_z dz} Q(+\infty) \, .  
\eeq
The difference with the $2$-flavor model considered in \cite{Shifman:2002jm}, and discussed in Section \ref{2flavors}, is that the two scalar fields at the edges of the Wilson line are now the same. This implies that  we do not have a strictly
massless gauge field localized on the wall. This is because the
expectation value of the matter field never exactly vanishes inside the
wall.

Let us denote the expectation value of the condensate $|Q|$
in the wall center $(z=z_0)$ by $v_0$,
\beq
v_0 = \left|
Q(z=z_0)\right|\,.
\label{veenot}
\eeq
In the limit in which $m\gg e v$,
the domain wall is thick and $v_0$ is very small.
A numerical fit in the range $0.3 \, v <m<0.6 \, v$ and $0.15<e<0.2$
shows that to a very good approximation 
\footnote{This formula, strictly speaking, is valid in the range of parameters
indicated above. However, it is likely that it works also for larger $d$, although
it is difficult to check this assumption because accurate numerical calculations 
are more difficult for larger values of $d$.}
\beq
 v_0 \approx v \exp \left(-0.88 \, \frac{m^2}{e^2 \, v^2}\right)=v \exp \left(-0.31 \, d \, m\right) \, . 
 \eeq 
The mass of the gauge field in the bulk is $ev$; Higgsing inside the wall is
exponentially weaker, so that the gauge field mass is $ \sim e \, v_0$.
  To what extent can we speak of
localization? The answer depends on the range of the parameters.

Since the domain wall is an object with thickness $d$  a low-energy
effective action makes sense only up to energy scales $\sim 1/d$.
At higher energies excitations of the wall
internal structure become important. Fluctuations of the wall as a whole in the
transverse direction  (Goldstone modes of the
translational symmetry) are massless. They always belong to the low-energy
effective action. Other --- massive --- excitations can be considered a
part of the  $2+1$ dimensional effective action as long as their mass is much
smaller than $1/d$. Physics changes in passing from one of the
following regimes to another:

$$ (i) \qquad
1/d \ll e \, v_0 \ll e \, v \,.$$
 In this regime the mass of the
gauge field inside the wall is much larger than $1/d$. We can not
speak about localization,   physics of the wall is essentially four-dimensional.

\vspace{1mm}

$$ (ii) \qquad
e \, v_0 \ll 1/d \ll  e \, v\,.$$
 This is the localization
limit. Up to energies $\sim 1/d$ the gauge field can be considered as a field localized on
the $2+1$ dimensional world volume.

\vspace{1mm}

$$ (iii) \qquad  e \, v_0 \ll e \, v \ll 1/d \,.$$ 
In this case the gauge field
is localized only up to energies $e \, v$. Due to leakage in the bulk
no localization occurs at energies
from $\sim ev$ to $1/d$.

\vspace{1mm}

Focusing on the regimes (ii) and (iii) we ask ourselves
whether or not a quasimoduli field lives on the wall in these cases.

\subsection{Modulus or quasimodulus on the wall}
\label{qmod}

First, we need to explain why we expect a U$(1)$ quasimodulus 
on the wall world volume.
We begin by presenting the simplest solutions that describe
localization of the gauge field: a constant magnetic field and a
constant electric field. From now on we will always work in the gauge
$A_z = 0$.

A constant magnetic field 
inside the wall is parallel to the wall surface. Let us assume  
the magnetic field to be aligned along the $x$ axis,
 $\vec{B}=B \hat{x}$ where $\hat{x}$ is the unit vector along $x$. 
We can construct it in the following
way. At negative $z$  we take the field $Q =v e^{i k y }$
and the gauge field $A_2 = k$ (or, which is the same, $A_y=-k$. 
At positive $z$ we take  $Q = v $ and
$A_y = 0$. In this way in  two vacua, to the left and to the right of
the wall,  the field configuration is  pure gauge. Inside the wall, $A_y$
linearly 
interpolates between $-k$ and $0$ on the interval of size $d$. 
The magnetic field is $B_x=
-\partial_z A_y = k/d$. The magnetic flux per unit length in the $y$ direction 
is $\int dz B_x = k$.
The magnetic field inside the wall is a vector on the wall since it can be
oriented either along $\hat{x}$ or along $\hat{y}$. 

The electric field inside the wall can only be perpendicular
to the wall surface, aligned along $\hat{z}$. Thus, on the wall it must be 
interpreted as a pseudoscalar.
To obtain such an electric field  inside the wall
consider $Q =ve^{i \omega t}$ at negative $z$ and $A_t =
\omega$. 
At positive $z$ we have $Q =v $  and $A_t =0$. 
Inside the wall $A_t$ linearly interpolates between $\omega$ and $0$.
The electric
field inside the wall is $E_z=-\partial_z A_t =\omega/d $.

Of course, from the $2+1$ dimensional point of view the picture must be dualized,
since in 2+1 dimensions it is the electric field $F_{0i}$ which is a vector while
the magnetic field $F_{12}$ is a scalar.
For example,
$E^{(2+1)}_x=B^{(2+1)}_x=-\partial_z A_x$ and
$B^{(2+1)}=B^{(2+1)}_z=-\partial_z A_t$. In other words,
\beq
F^{(2+1)}_{\mu \nu}=\frac12 
\epsilon^{\mu \nu\rho z} F_{\rho z}\,,\qquad \mu ,\nu , \rho = 0,1,2\,.
\label{dual1}
\eeq
Now we can further dualize $F^{(2+1)}_{\mu \nu}$ \`a la Polyakov,
expressing $F^{(2+1)}_{\mu \nu}$ in terms of a phase fields $\sigma$,
\beq 
\frac{1}{2} \epsilon_{\mu\nu\rho} F^{(2+1)}_{\nu\rho} = \frac
1d
\partial_{\mu} \sigma 
\label{sigm}
\eeq
Assembling everything together we have, with our gauge choice  
\bea 
 B^{(3+1)}_x =-\partial_z A_y \qquad &=& \qquad E^{(2+1)}_{x}
  =\frac1d \, \partial_y \sigma \,,
 \\[2mm]
 B^{(3+1)}_y=\partial_z A_x \qquad &=& \qquad  E^{(2+1)}_{y} 
 = -\frac1d \,  \partial_x \sigma \,,
 \\ [2mm] 
E^{(3+1)}_z=-\partial_z A_t \qquad &=& \qquad   B^{(2+1)}
= \frac1d \,  \partial_t \sigma \,.
\eea  
Note
that the angle field $\sigma$ exactly corresponds  to the
phase of $Q $ at negative $z$  relative  to that at  positive $z$.

We can use unitary gauge which ensures that $Q=v$ in both
vacua at $z\to\pm\infty$. In this gauge the constant magnetic
field inside the wall looks as follows. The gauge field $A_\mu=0$ in 
both vacua, while inside the wall
\beq
A_z=\frac1d\,\sigma(x,y),\qquad A_x=A_y=0.
\label{unitarygauge}
\eeq
This gauge potential gives the magnetic field strength
shown in equations above.

As discussed before, one can consider the gauge invariant order parameter (\ref{order}). The kinetic term for the gauge field, expressed as a function of the
$\sigma$ modulus, takes the form (cf. \cite{Shifman:2002jm})
\beq {\cal L}_{2+1}= \frac{1}{2\,e^2_{3+1}\,d}\,\, 
\partial_{\mu}\sigma\partial_{\mu}\sigma \, .
\label{kint}
\eeq

\subsection{Potential for the quasimodulus}
\label{potqm}

As was mentioned, in the model at hand, 
the phase field $\sigma$ is not an exact modulus. A
potential $V(\sigma)$ is generated forcing $\sigma=0$ in the true
vacuum which is unique. 
In the localization limit of large $d$ we can nevertheless speak of a
$2+1$ dimensional effective theory for $\sigma$, since the $\sigma$ field  mass is much
smaller than the excitation energy of the domain wall $\sim 1/d$. In the leading approximation
the $\sigma$ field Lagrangian will be of the sine-Gordon type,
 \beq
\label{effectiveaction} 
{\cal L}_{2+1}= \frac{1}{2e^2\,d}\,\,
\partial_{\mu}\sigma\partial_{\mu}\sigma - 
\Delta \cdot\left(\sin{\frac{\sigma}{2}}\right)^2 
\eeq 
where $\Delta$ is the difference between two tensions: the tension of the
$\sigma=0$ wall and and that of the $\sigma=\pi$ wall. The mass of $\sigma$ is
\beq 
\label{guess} 
 m_{\sigma} \approx \sqrt{ \frac  {  \Delta \, e^2 \, d}{2}} \, .
\eeq

Explicit calculations of the quasimodulus potential $V(\sigma)$
are not easy. The minimal tension domain wall
corresponds to the lowest energy $\sigma=0$. Walls with
$\sigma \neq 0$ are not static solutions of the equation of motion
since they decay.
To study such solutions of the equation of motion we  have to deal
with more than one parameter.

\begin{figure}[h!tb]
\epsfxsize=11cm \centerline{\epsfbox{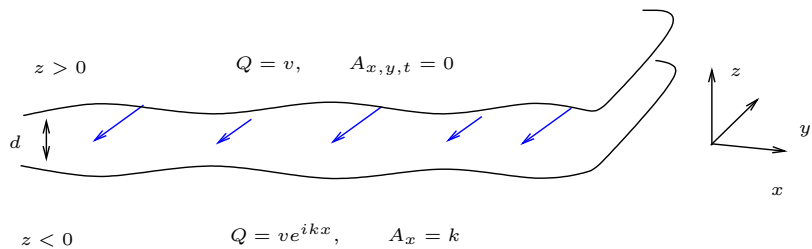}}
\caption{{\protect\small An (almost) constant magnetic 
field inside the domain wall. From the point of view of the quasi modulus $\sigma$ 
it correspond to an ``almost'' linear solution $\sigma =kx + \dots$. 
The dots stand for small modulations.
When $\sigma \approx 0$, the wall is thinner and the magnetic field larger. At $\sigma \approx \pi$ the magnetic field is smaller and the wall thicker. The amplitude of the oscillations in $|\vec{B}|$ and thickness depends on the mass of the quasimodulus. In the limit of vanishing
 mass we recover the constant magnetic field solution.}} 
 \label{constantb}
\end{figure}

To explain what the last sentence means
we consider the solution with a constant magnetic
field in the $\hat{y}$ direction (see Fig.~\ref{constantb}).

As a first approximation, we can choose $Q=ve^{ikx}$ and $A_x =k$ at
$z<0$, and $Q=v$ and $A_x=0$ at $z>0$, similar to the discussion above. This
gives a constant magnetic field $B_y = k/d$ inside the wall. But it
is clear that, due to a nonvanishing (albeit small) $Q$ condensate inside
the wall, this is {\em not} the exact solution. The latter requires
modulations of $k$. One can
understand this circumstance both from the bulk and from the brane point of view. 
The bulk explanation (see Fig.~\ref{constantb}) is as follows. Due to
topological reasons, the  field $Q$ must exactly vanish at some $x$ and $z\approx 0$ each
time the relative phase $\sigma$ rotates by $2\pi$. 
The lines (in the $y$ direction) on which $Q$
vanishes are the lines where the magnetic field reaches its
maximum. We also expect the thickness of the wall to be a little bit
larger around these lines. 

From the point of view
of the $2+1$ dimensional effective action (\ref{effectiveaction}) it is also clear
that  $\sigma(y) = k x$ with $k$ constant is not a solution
once the sine term is switched on. The
derivative $d\sigma/dx$ will be larger at the top of the
potential ($\sigma =\pi$) and smaller at the bottom ($\sigma=0$).

\begin{figure}[h!t]
\begin{center}
$\begin{array}{c@{\hspace{.2in}}c} \epsfxsize=2.5in
\epsffile{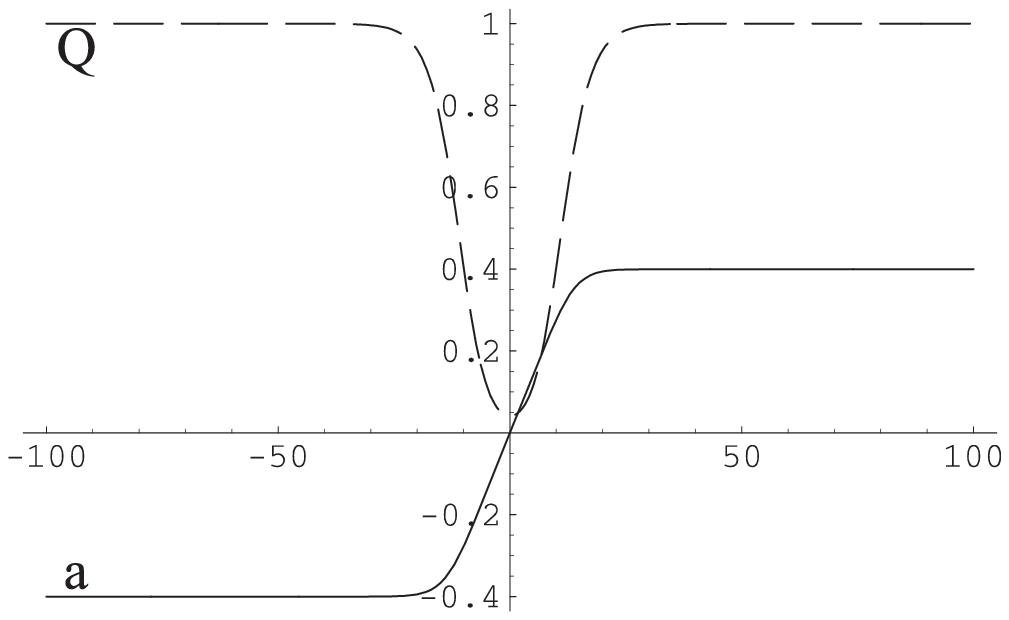} &
    \epsfxsize=2.5in
    \epsffile{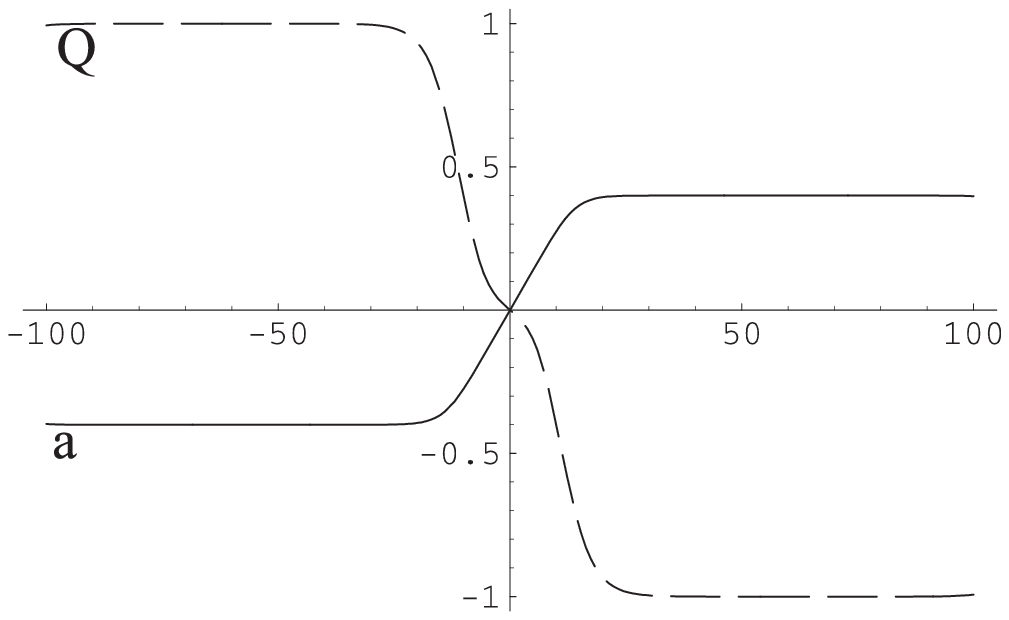}
\end{array}$
\end{center}
\caption{\footnotesize Left: the stable wall ($\sigma =0$).
  Right: the unstable wall ($\sigma =\pi$). In this specific example, even though
the Higgs VEV is not so small in the middle of the stable wall, there is
  just a $2\%$ t difference between the tensions of the unstable and stable walls. }
\label{stableandunstable}
\end{figure}
\begin{figure}[h!t]
\begin{center}
$\begin{array}{c@{\hspace{.2in}}c} \epsfxsize=2.5in
\epsffile{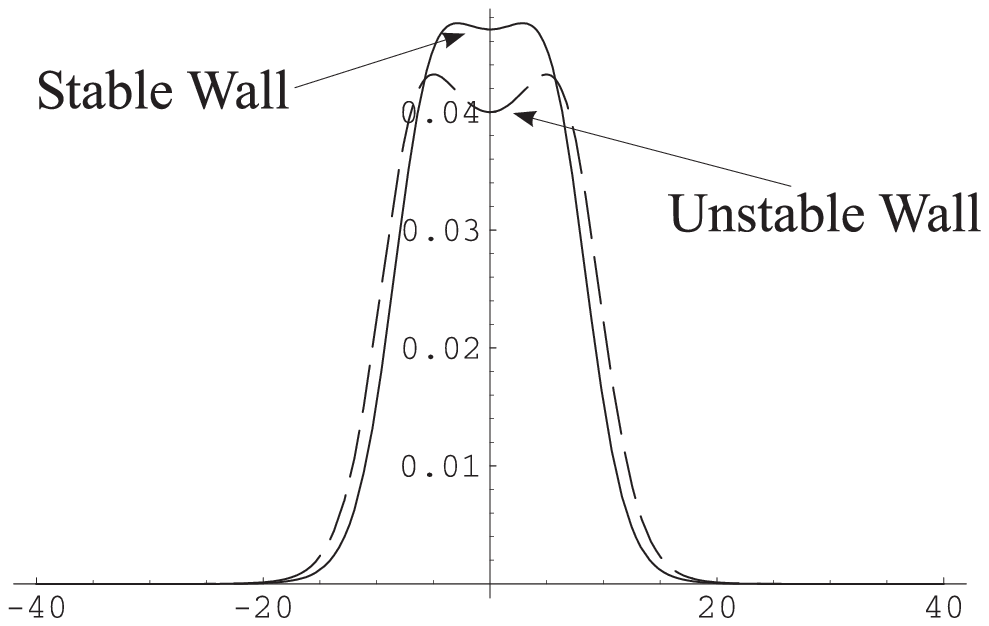} &
    \epsfxsize=2.5in
    \epsffile{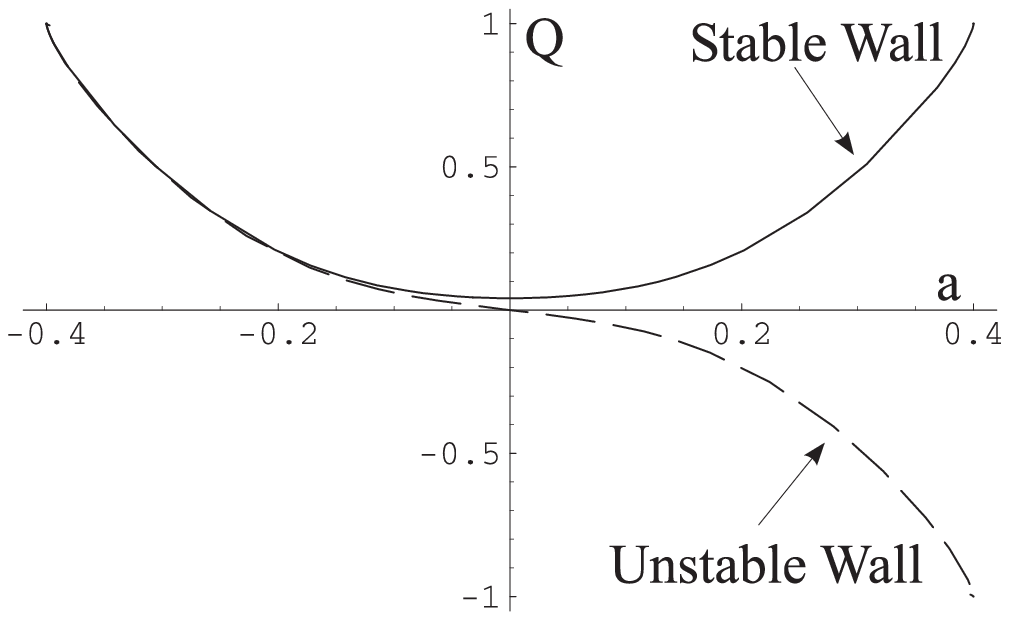}
\end{array}$
\end{center}
\caption{\footnotesize Left: the energy density inside the stable wall (solid line) and the
unstable wall (dashed line). In the center of the wall, the unstable
wall has a lower energy density. On the other hand, the unstable wall
is thicker and this make its total tension larger than that of the stable wall. 
Right: both domain walls in the $(a,Q)$ plane.}
\label{densi}
\end{figure}

There is a trick one can use to detect the quasimodulus. At $\sigma =\pi$ the wall solution
corresponds to the
maximum of the potential $V(\sigma)$. It is stationary but unstable.
In this solution the $Q$ field vanishes exactly in the middle of the wall, at $z=0$.
We can easily get the solution if we impose two conditions:
(i) the field $Q$ is real; (ii)
$Q$ changes sign in passing from one side of the wall to the other, see Figs.
\ref{stableandunstable} and \ref{densi}. The difference between 
the tensions of the unstable and stable walls
will determine $\Delta$ in Eq.~(\ref{effectiveaction}).

The mass term for the $\sigma$ field, is what induces confinement inside the domain wall. 
In Figure \ref{constantb} we discussed how a constant magnetic field solution is deformed by the mass term. 
There are modulations of the thickness of the wall, and magnitude of the magnetic field. 
These modulations are nothing but the flux tubes, all lined together. 
A single winding in the phase $\sigma$,  a kink ot the sine-Gordon model (\ref{effectiveaction}), is a domain line from the wall point of view. 
The mass term is what stabilizes the thickness of this kink, that otherwise would spread to $\infty$. 
These, from the point of view of the bulk theory, are the confining vortices bounded inside the domain wall.

\subsection{A direct estimate of the $\sigma$ mass term}

The Lagrangian of the model under consideration has no exact or approximate
global U(1) symmetry. However, the domain wall solution does have such an 
approximate symmetry. Indeed, let us start from the limit $m/(ev)\to\infty$.
In this limit the wall edges become infinitely sharp while the absolute
value of the field $Q$ inside the wall vanishes. Then the field configuration under consideration has two U(1) symmetries: $Q(z>z_0+d/2)\to e^{i\alpha}Q(z>z_0+d/2)$ and 
$Q(z<z_0- d/2)\to e^{i\beta}Q(z <z_0-  d/2)$, with independent phases $\alpha$ and $\beta$.
The common U(1) symmetry, with $\alpha =\beta$, is gauged.
The U(1) phase rotation of $Q(z> z_0+ d/2)$ relative to $Q(z<z_0- d/2)$
remains as an exact global U(1) symmetry.

If we keep $m/(ev)$ finite, then $Q(z_0)$ does not vanish and, because of continuity of the field
$Q$, the above symmetry becomes approximate. 
This is due to the fact that now one must continuously evolve the  phase of $Q$
from zero at $ z=z_0- d/2$ up to a given $\sigma$ at $ z=z_0+ d/2$.
With the $z$ dependent phase inside the wall, the wall configuration  no longer presents the exact solution. 
The question is how to deal with such a situation.

This is quite similar to the treatment of quasizero modes
in the instanton-anti-instanton background field.
To deal with them one routinely  uses the valley method 
\cite{BY}: in the functional space
one finds a direction corresponding to the bottom of the valley
and fixes the coordinate along this direction ``by hand." 
In the problem at hand the coordinate to be fixed can be defined
as the phase $\sigma$ in the expression $Q(z=z_0+ d/2)
= e^{i\sigma} Q(z=z_0- d/2)$  (in the $A_z=0$ gauge).
For simplicity we will assume $\sigma(z=z_0+d/2)$ to be small. 
Then, given the above boundary condition, minimization of the energy functional
gives that inside the wall 
\beq
\sigma (z) \approx \frac{z-z_0+(\delta/2)}{\delta}\, \sigma\,,
\label{28}
\eeq
where $\delta$ is the size of the region inside the wall where phase $\sigma$
is changing, $\delta < d$. It is a free parameter and we estimate it below
by minimization procedure.

As for the ansatz for the scalar  field inside the wall we choose the 
 sum of the profile functions (\ref{Qlprofile}) and (\ref{Qrprofile})
at the left and right edges of the wall.  We have
\beq
Q\approx v\,\left[e^{-\frac{2m}{d}(z-z_0+\frac{d}{2})^2} +
e^{-\frac{2m}{d}(z-z_0-\frac{d}{2})^2}\right]\,
\exp{\left(i\,\frac{z-z_0+(\delta/2)}{\delta}\, \sigma\right)}.
\label{Qinside}
\eeq

It is quite obvious that the only extra term in the energy functional (\ref{lagr}),
(\ref{potep}) comes from $|\partial_z Q |^2$ with the derivative acting on the phase,
i.e. $(Q \partial_z \sigma )^2$. Thus, the potential energy associated with 
$\sigma \neq 0$ (at small $\sigma$)
is
\beq
V(\sigma ) = \int dz\, \left( Q\frac{\sigma}{\delta}\right)^2 \sim \sigma^2\,
\frac{v^2}{\delta^2}\,\int_{-\delta/2}^{\infty} dz \,e^{-\frac{2m}{d}(z+\frac{d}{2})^2}
\sim  \sigma^2\,\frac{v^2}{\delta^2 ev}\,e^{-\frac{m}{2d}(d-\delta)^2}   \,.
\label{sigmapotint}
\eeq

Minimizing this with respect to $\delta$ we get
\beq
e^2v^2(d-\delta)\sim \frac{1}{\delta}.
\eeq
Assuming that $\delta\ll d$ we have
\beq
\delta\sim \frac1{e^2v^2\,d}\sim \frac1m,
\label{Delta}
\eeq
which confirms our assumption, since $\delta/d\sim \frac{e^2 v^2}{m^2}\ll 1$.

Substituting $\delta$ from (\ref{Delta}) to the potential (\ref{sigmapotint}) we 
get
\beq
V(\sigma) \sim \sigma^2\,v^2\,\frac{m^2}{ev}\,e^{-\frac{md}{2}},
\eeq
which shows that the potential for $\sigma$ is exponentially small and is determined
by the value of the scalar field  in the middle of the wall. Using more accurate 
numerical
estimate (\ref{veenot}) for the value $v_0$ we finally get
\beq
V(\sigma )  \sim \frac{v_0^2\,m^2}{ev}\, \sigma^2\,.
\eeq
 This estimate implies, in turn, that
\beq
\tilde{m}_\sigma 
\sim ev_0\,\left(\frac{m}{ev}\right)^{3/2} \, . 
\label{29}
\eeq

\subsection{A numerical check}

We have now two different estimations of $m_{\sigma}$.
One of them (Eq.~(\ref{guess})) is given in term of $\Delta$,
which is the difference in tension between the 
 $\sigma=\pi$ and the $\sigma=0$ wall.
 The other one (Eq.~(\ref{29})) is given in terms of the 
 residual condensate $v_0$ at the center of the domain wall.
It is interesting to compare them for a cross check.

\begin{table}[h]     
\begin{center}    
\begin{tabular}{|c|c|c|c|c|c|c|}
\hline
 \multicolumn{2}{|c|}{} & \multicolumn{5}{|c|}{m} \\ \cline{3-7}
  \multicolumn{2}{|c|}{} &$0.30$ &$0.35$ &$0.40$ & $0.45 $& $0.50 $ \\ \hline
  \multirow{5}{*}{e} & $0.2$ &  $2.5\times 10^{-2}$ & $6.0\times 10^{-3}$ & $1.1\times 10^{-3}$ & $1.6\times
   10^{-4}$ & $1.8\times 10^{-5}$ \\ \cline{2-7}
  & $0.19$& $1.6\times 10^{-2}$ & $3.2\times 10^{-3}$ & $4.8\times 10^{-4}$ & $5.7\times
   10^{-5}$ & $5.1\times 10^{-6}$ \\ \cline{2-7}
    & $0.18$&  $  9.0\times 10^{-3}$ & $1.5\times 10^{-3} $& $1.8\times 10^{-4}$ &$ 1.7\times
   10^{-5}$ & $1.1\times 10^{-6} $\\  \cline{2-7}
      & $0.17$& $ 4.7\times 10^{-3}$ & $6.2\times 10^{-4}$ & $5.8\times 10^{-5}$ & $3.9 \times
   10^{-6}$ & $1.9\times 10^{-7} $\\   \cline{2-7}
        & $0.16$&  $2.1\times 10^{-3} $&$ 2.1\times 10^{-4}$ & $1.5\times 10^{-5}$ & $7.0\times
   10^{-7} $&$ 2.2\times 10^{-8}$ \\   \cline{2-7}
\hline
\end{tabular}
\caption{\footnotesize Values of $\Delta$ for $v=1$ and
different values of $(m,e)$.}     
\label{delta}  
\end{center}
\end{table} 
 
\begin{table}[h]     
\begin{center}    
\begin{tabular}{|c|c|c|c|c|c|c|}
\hline
 \multicolumn{2}{|c|}{} & \multicolumn{5}{|c|}{m} \\ \cline{3-7}
  \multicolumn{2}{|c|}{} & $0.30$ &$0.35$ &$0.40$ & $0.45 $& $0.50 $ \\ \hline
  \multirow{5}{*}{e} & $0.2$ &  $2.6\times 10^{-1}$ & $1.1\times 10^{-1}$ & $4.1\times 10^{-2}$ & $1.4\times
   10^{-2}$ & $4.5\times 10^{-3}$ \\ \cline{2-7}
  & $0.19$& $2.0\times 10^{-1}$ & $7.7\times 10^{-2}$ & $2.7\times 10^{-2}$ & $8.4\times
   10^{-3}$ & $2.3\times 10^{-3}$ \\ \cline{2-7}
    & $0.18$&  $1.5\times 10^{-1}$ & $5.2\times 10^{-2}$ & $1.6\times 10^{-2}$ & $4.5\times
   10^{-3}$ & $1.1\times 10^{-3} $ \\   \cline{2-7}
      & $0.17$&   $1.0\times 10^{-1} $& $3.2\times 10^{-2}$ &$ 9.0\times 10^{-3} $& $2.2\times
   10^{-3} $&$ 4.5\times 10^{-4}$ \\   \cline{2-7}
        & $0.16$& $6.8\times 10^{-2}$ & $1.9\times 10^{-2}$ & $4.5\times 10^{-3} $& $9.1\times
   10^{-4}$ & $1.5\times 10^{-4} $ \\ \cline{2-7}
\hline
\end{tabular}
\caption{\footnotesize Values of $ v_0 $ for $v=1$ and
different values of $(m,e)$.}     
\label{condensato}  
\end{center}
\end{table}

In Table \ref{delta}    numerical results for $\Delta$ are shown
for some values of the parameters. In Table \ref{condensato}
the corresponding values of $v_0$ are presented.
We can then use these values for a comparison the two independent
estimations $m_\sigma$, $\tilde{m}_{\sigma}$.
The results are shown in Table \ref{testnum}. The proximity of the ratio $m_\sigma/\tilde m_\sigma$ to unity is obvious.
The agreement is indeed quite good.
We can also verify that the localization condition $e \, v_0 \ll 1/d$ is satisfied
very well for $m \geq 0.4 \, v$ and $e<0.2$.

\begin{table}[h]     
\begin{center}    
\begin{tabular}{|c|c|c|c|c|c|c|}
\hline
 \multicolumn{2}{|c|}{} & \multicolumn{5}{|c|}{m} \\ \cline{3-7}
  \multicolumn{2}{|c|}{} & $0.30$ &$0.35$ &$0.40$ & $0.45 $& $0.50 $ \\ \hline
  \multirow{5}{*}{e} & $0.2$ &  $1.08$ & $1.08$ & $1.08$ & $1.05$ & $1.01$ \\ \cline{2-7}
  & $0.19$& $1.07$ & $1.08$ & $1.07$ & $1.03$ & $0.99$ \\ \cline{2-7}
    & $0.18$&  $1.08$ & $1.08$ & $ 1.05$ & $1.01$ & $0.97$ \\   \cline{2-7}
      & $0.17$&   $1.08$ & $1.07$ & $1.03$ & $0.99$ & $0.95$ \\   \cline{2-7}
        & $0.16$& $1.08$ & $1.06$ & $1.01$ & $0.97$ & $0.92$ \\ \cline{2-7}
\hline
\end{tabular}
\caption{\footnotesize Values of $m_{\sigma}/\tilde{m}_{\sigma}$ for $v=1$ and
different values of $(m,e)$.
$m_\sigma$ is given in Eq.~(\ref{guess}) and 
$\tilde{m}_{\sigma}$ given by Eq.~(\ref{29}).
}     
\label{testnum}  
\end{center}
\end{table}

\section{A Modified Seiberg--Witten framework}
\label{seibergwitten}

\subsection{Theoretical Setting}
\label{these}

We now turn to 
a strong coupling example of the confinement
phenomenon inside domain walls. 
The model we will dwell on below is a rather straightforward modification of the
Seiberg--Witten (SW) model \cite{Seiberg:1994rs}, which supports both domain walls and strings.

The theory of interest is $\mathcal{N}=2$ gauge theory, 
with the gauge group ${\rm U}(2)= {\rm SU}(2) \times {\rm U}(1) /Z_2 $,
with no matter hypermultiplets. The following superpotential which breaks
the extended supersymmetry down to $\mathcal{N}=1$  is then added: 
\beq W= \alpha \,  \Tr \, \left(
\frac{\Phi^3}{3} - \xi \Phi \right). \label{superpotential} \eeq Classically, we have
three vacua, with $\phi$ equal to
\beq  
\label{classicalvacua}
\left( \begin{array}{cc} \phantom{.}\sqrt{\xi} & 0\\[2mm]0  & \phantom{-}\sqrt{\xi} 
\end{array}\right) \ , \qquad 
\left( \begin{array}{cc} \phantom{.}\sqrt{\xi} & 0 \\[2mm]
0 & -\sqrt{\xi} \end{array}\right) \ , 
\qquad 
\left( \begin{array}{cc}- \sqrt{\xi} & 0\\[2mm]
0 &- \sqrt{\xi} \end{array}\right) \ .
   \label{threevacua}
\eeq   
The first and the last vacua preserve the non-Abelian SU(2) gauge symmetry. Strong coupling effects
\`{a} la Seiberg and Witten will then split each of them into two vacua (the monopole 
and dyon vacua). The vacuum in the middle preserves only the U$(1) \times {\rm U}(1)$ gauge symmetry,
and is not split. We, thus, expect in total  five vacua, for   generic values of $\xi$. 
   
If we set $\alpha$ at zero, the U$(1)$ and  SU$(2)$ sectors get completely
decoupled. Since there are no matter hypermultiplets, only
a nonvanishing superpotential can make the two sectors communicate with
each other.  
Dynamics of the U$(1)$ sector is trivial, while the
SU$(2)$ sector is described by the Seiberg--Witten  solution \cite{Seiberg:1994rs}. We parametrize 
the  moduli space by 
\beq \Phi =
\left(\begin{array}{cc}
a_0 + a_3  & 0 \\
0 & a_0 - a_3 \\
\end{array}\right).
\label{pms}
\eeq
The conventional SW solution is written in terms of the invariant
\beq
 u= 2 a_3^2 \, .
 \label{swi}
 \eeq
In our case
 \beq {\rm Tr}\, \Phi=2 a_0,
\qquad  {\rm Tr}\, \Phi^2=2 a_0^2 + 2 a_3^2, \qquad   {\rm Tr}\,
\Phi^3=2 a_0^3 + 6 a_3^2 a_0 \,,
\label{fpop}
\eeq
and there is an invariant way to parameterize the moduli space 
\cite{Cachazo:2002ry,Cachazo:2002zk} by
\beq u_1 = \Tr\, \Phi\, , \qquad u_2 = \frac{1}{2} \Tr\,  \Phi^2  \,.
\label{awfms}
\eeq
 The trace of  of $\Phi^3$ can be expressed in terms of $u_{1,2}$,
\beq
\Tr \, \Phi^3 = \frac{3}{2} (\Tr \, \Phi) (\Tr\,  \Phi^2)-\frac{1}{2} (\Tr \, \Phi)^3 \, . 
\eeq 
This relation is not modified by quantum corrections. 

After the superpotential (\ref{superpotential}) is switched on, the moduli space is lifted.
Five discrete vacua  described above survive.  
This is a special case of the  set-up considered in Refs.~\cite{Cachazo:2002ry,Cachazo:2002zk}.
The positions of the vacua  are the following (see Appendix \ref{vacua} for more details). 
The value of $u_2$ is  $\xi$ for all  five vacua. It is not modified by quantum corrections.  
The Coulomb vacuum in the middle is not modified by quantum correction either.
It lies at 
\beq 
u_1=0\,, \quad u_2 = \xi \,.
\label{miva}
\eeq 
The monopole-$1$ and dyon-$1$ vacua  $\phi={\rm diag}\,(\xi,\xi)$ are at 
\beq
 u_1 = 2\sqrt{\xi - \Lambda^2 }, \,\,\, u_2=\xi \quad{\rm and} \quad
  u_1 = 2\sqrt{\xi + \Lambda^2 }, \,\,\, u_2=\xi \,,
\label{va1}
\eeq
respectively. The dyon-$2$ and monopole-$2$ vacua from $\phi={\rm diag}\, (-\xi,-\xi)$ are at 
\beq
u_1 = -2\sqrt{\xi - \Lambda^2 }, \,\,\, u_2=\xi \quad{\rm and} \quad
  u_1 = - 2\sqrt{\xi + \Lambda^2 }, \,\,\, u_2=\xi \,.
\label{va2}
\eeq
\begin{figure}[h!t]
\epsfxsize=8.5cm \centerline{\epsfbox{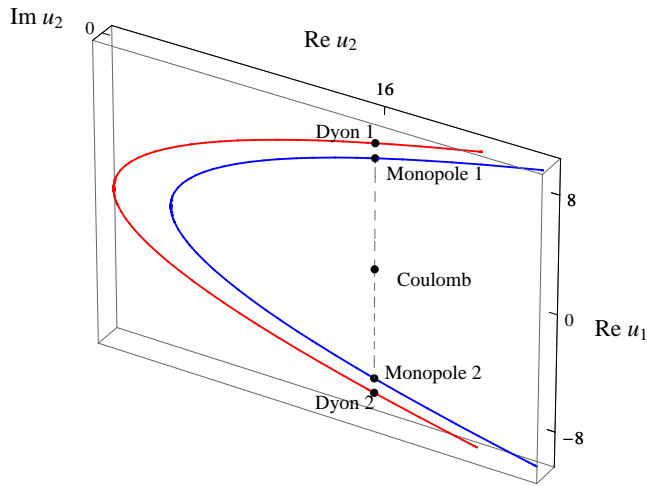}}
\caption{{\protect\small  Five vacua of the model
in the limit $\sqrt{\xi} \gg \Lambda$. (In the plot we set $\Lambda=1$, $\xi=4$). The dashed line connects the monopole-$1$, Coulomb, and monopole-$2$ vacua. This is the composite domain wall we will analyze in what follows. Note that all five vacua are aligned.  }} 
\label{vacuaxibig}
\end{figure}
\begin{figure}[h!b]
\epsfxsize=7cm  \centerline{\epsfbox{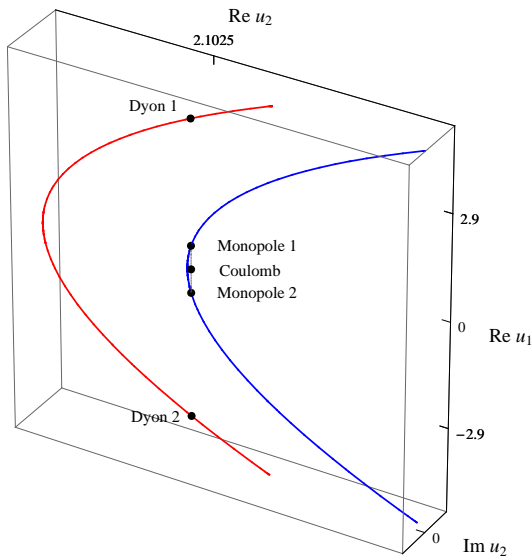}}
\caption{\footnotesize  As $\xi$ becomes smaller, we reach a critical point where the Coulomb vacua lies exactly on the monopole singularity. Around this point we can use the magnetic effective action to describe the wall. It remains weakly coupled in all three vacua of interest ($\Lambda=1$, $\xi=1.45$ in this plot).}
\label{vacuaxismall}
\end{figure}
In the limit $\sqrt{\xi} \gg \Lambda$ the Coulomb vacuum is such that the electric coupling is small,
\beq 
e_{3}^2 = \frac{1}{\ln\,
(\sqrt{\xi}/\Lambda)}\,.
\label{eccs}
\eeq
The five vacua in this limit are depicted in Fig.~\ref{vacuaxibig}.

As $\xi$ decreases and becomes of order $\Lambda$,  the Coulomb vacuum enters a strong coupling regime. At the critical value  $\xi =  \Lambda^2 $  the Coulomb vacuum lies exactly in the monopole singularity and coalesces with  two monopole vacua.  Around this critical value, the  Coulomb vacuum is such that the magnetic coupling is small, so we can use the same set of low-energy effective variables
to describe both the Coulomb and the confining vacua.
In this section we will study (in the limit   $|\xi -  \Lambda^2| \ll \Lambda^2 $) the domain walls connecting 
the Coulomb and the confining vacua. It is  convenient to introduce   a dimensionless parameter
 \beq
\epsilon^2=\frac{\xi-\Lambda^2}{\Lambda^2}\,. 
\label{dip}
\eeq
We will work in the limit $|\epsilon|\ll 1$, but keeping
three vacua separate (i.e. $\epsilon \neq 0$).

The effect of the superpotential $W$ in the infrared theory, near the monopole vacuum,
is described by the following effective superpotential: 
\beq
\tilde{W}= \frac{A_{3D} M \tilde{M}}{\sqrt{2}} + \frac{ \alpha  (2
A_0^3+6 u(A_{3D}) A_0 ) }{3} - 2 \alpha \xi A_0 \, , 
\label{effsu}
\eeq 
where we
can use
 \beq u(A_{3D}) \approx \Lambda^2-2 i \Lambda
A_{3D}-\frac{1}{4} A_{3D}^2+\ldots \,.
\eeq
 The $F$-term part of the scalar potential  is
\beq 
V_F= e_{3D}^2|\frac{ M \tilde{M}}{\sqrt{2}}+ 2 \alpha u'
A_0|^2+ 4 e_0^2 \alpha^2 |  A_0^2 +  u   - \xi |^2  + \frac{| M
A_{3D}|^2+|\tilde{M} A_{3D}|^2}{2} \,,
\label{ftpa}
\eeq 
while the $D$-term part  is
\beq
V_D=\frac{e_{3D}^2}{2} (M M^\dagger - \tilde{M}
\tilde{M}^\dagger)^2
\,,
\label{dtpa}
\eeq
where $M$, $\tilde M$ are the monopole superfields.
The vacuum expectation values in the confining vacua are (in the limit  $\epsilon\ll 1$)
 \beq
A_{3D}=0, \qquad A_0=\pm \Lambda \epsilon, \qquad M \tilde{M}=  \pm
4 \sqrt{2}\, i\,  \Lambda^2 \alpha \epsilon  \,. \eeq
The dual description is valid in
these vacua
provided that the monopole condensate is $\ll \Lambda^2$ implying
$|\alpha \epsilon |\ll 1$.  

The Coulomb vacuum is defined by the following constraints: 
\beq  
 M = \tilde{M} =0 \, , \qquad
A_0=0\, , \qquad u(A_{3D})= \xi\, . 
\label{cvc}
\eeq 
The VEV of $A_{3D}$
in the limit $\epsilon\ll 1$ is 
 \beq 
 A_{3D}=\frac{\epsilon^2 \Lambda
i}{2} \,. 
\label{cvcp}
\eeq 

A nice feature of the limit $\epsilon\ll1$ is that we can use
the same weakly coupled effective description in all three vacua
of interest (Coulomb, monopole-1 and monopole-2). This unified
description is also valid for the domain wall interpolating between them.
In this sense the situation drastically differs from the case considered in 
Ref.~\cite{KSY},
where no unified description  was possible for the domain wall interpolating
between the monopole and the dyon vacua considered in \cite{KSY}.

The kinetic terms are 
 \beqn
\mathcal{L}_{\rm kin}
&=&\frac{1}{4e_{3D}^2} \left(F_{3D}^{\mu \nu}\right)^2
+\frac{1}{4e_{0}^2} \left(F_{0}^{\mu \nu}\right)^2
\nonumber\\[3mm]
&+&
 \frac{1}{2e_{3D}^2}
(\partial_\mu {a_{3D}})^2+ \frac{1}{2e_{0}^2} (\partial_\mu
{a_{0}})^2+ |\nabla_\mu M|^2+ |\nabla_\mu \tilde{M} |^2 \,. 
\eeqn
In the confining vacua the value of the coupling $e_{3D}$
is determined by the monopole condensate,
 \beq e_{3D}^2 = \frac{1}{\ln\,
(\Lambda/\sqrt{|M \tilde{M}|})} \approx \frac{1}{\ln\, (
1/\sqrt{| \alpha \epsilon |} )}\,.
\label{e3d}
 \eeq 
 In the Coulomb vacuum the expression for
$e_{3D}^2$  can be found from the Seiberg--Witten expression 
for $\tau=\theta/(2 \pi)+ 4 \pi i/g^2 $,
\beq 
e_{3D}^2  \approx \frac{1}{\ln \,(
1/ |\epsilon |  )}\,. 
\label{e3dp}
\eeq 
An effective coupling $e_{3D}(z)$ along the domain wall profile
is a function of the field condensates inside the wall. It interpolates
between the two values above, (\ref{e3d}) and (\ref{e3dp}). 
If we choose $\alpha \sim \epsilon$, then $e_{3D}^2$
is approximately constant.

Replacing the leading terms in $u(A_{3D})$ in the scalar potential, we get
\beqn 
V 
&=&
 e_{3D}^2 \left|\frac{ M \tilde{M}}{\sqrt{2}}-4 i \alpha \Lambda
A_0\right|^2 + 4 e_0^2  \alpha^2 \left|  A_0^2 -2 i \Lambda A_{3D} + \Lambda^2
- \xi \right|^2 \nonumber\\[3mm]
& +&
\frac{ | M A_{3D}|^2+|\tilde{M} A_{3D}|^2 }{2}
+\frac{e_{3D}^2}{2} (M M^\dagger - \tilde{M} \tilde{M}^\dagger)^2
\,.
\label{vscal}
\eeqn

\subsection{Elementary and composite walls}
\label{eacw}

We look for the wall solution in  the ansatz: 
\beq
 |\tilde{M}|=  |M |\,
,
\label{ansa}
\eeq
which automatically guarantees vanishing of the 
$D$ term. 
We will assume
the phases of $\tilde{M}$ and   $M$ to be constant constant 
in our wall solution.
This assumption is to be checked {\em a posteriori}.
Once the phases are constant, they can be chosen at will by virtue of
a global gauge rotation. 
It is convenient to choose
 \beq
\tilde{M}=i M\,. 
\label{coph}
\eeq
The value of the superpotential in the two confining vacua,
monopole-1 and monopole-2,
is
\beq
W = \pm\, \frac{4}{3}\, \Lambda^3 \,\alpha \epsilon^{3} \, . 
\label{vccv}
\eeq 
In the
Coulomb vacuum the superpotential vanishes. Hence,  the
tension of the BPS wall interpolating between monopole-1 and monopole-2 vacua,
if it existed,  would be
twice the tension of the BPS wall interpolating between the Coulomb
and  confining vacua. The latter walls will be referred to as elementary.
The former wall can be called composite.

\begin{figure}[h!tb]
\epsfxsize=7cm \centerline{\epsfbox{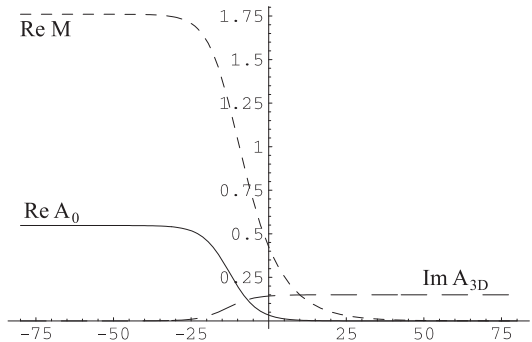}}
\caption{{\protect\small Elementary wall at $\mu=0$ (see Eq.~(\ref{modsp})), $A_0$ (solid), $A_3$ (long dashes)
 and $M$ (short dashes). }} 
 \label{single}
\end{figure}

For the elementary wall, we can take $M$ and $A_0$ real 
and $A_{3D}$ pure
imaginary. Then we can write the following BPS equations:\,\footnote{ 
Note that $e_{3D}$ is a function of the fields $M \tilde{M},\,\, a_{3D}$;
this does not change the form of the BPS equations.
Our choice $\alpha \approx \epsilon$ guarantees   
constancy of $e_{3D}$ to a very good approximation.
For simplicity, in the numerical calculations we keep $e_{3D}$ constant.}
 \beqn 
&& \frac{\partial M}{\partial z}- i \frac{M \, A_{3D}}{ \sqrt{2}} =0 \,,
\nonumber\\[3mm]
 && \frac{\partial A_0}{\partial z}-2 \alpha e_0^2 (A_0^2 - 
2 i \Lambda A_{3D}-\xi+\Lambda^2)=0 \, ,
\nonumber\\[3mm]
&& \frac{\partial A_{3D}}{\partial z}+i e_{3D}^2
\left( \frac{M^2}{\sqrt{2}} - 4 \alpha \Lambda A_{0}\right) =0 \, .
\label{eq43}
\eeqn
The profile functions for the elementary wall interpolating between the
confining and the Coulomb vacua, are shown if Fig.~\ref{single}.
In this theoretical set-up, no BPS wall interpolating between  two
confining vacua, monopole-1 and monopole-2,    exists. In other words, a composite wall
built of two elementary walls at a finite distance from each other,
does not exist.
Supersymmetric solutions correspond to viscous flows in the first-order equations,
starting from monopole-1, following the profile $\tilde W$ and ending in
monopole-2.
It is not difficult to see that such flow cannot be realized in the case at hand.
A field configuration interpolating between monopole-1 and monopole-2 is
always time-dependent; it represents two elementary walls moving 
under the influence of a repulsive 
force between them (see Ref.~\cite{Portugues:2001ah}
for a discussion  in supersymmetric sigma models). 
This force falls off exponentially with the wall separation.

\subsection{Composite wall stabilization}
\label{cws}

In order to avoid the problem discussed
in Sect.~\ref{eacw} and stabilize the composite domain wall, an extra term  is
introduced in the superpotential,
\beq 
W=  \alpha \left( {\rm Tr}  \left(
\frac{\Phi^3}{3} - \xi \Phi \right  ) + \frac{i \mu}{2} ({\rm
Tr} \,\Phi)^2 \right),
\label{modsp}
\eeq
where $\xi$ is chosen as a real parameter, with $\xi> \Lambda^2$,  
and $\mu$ is a real mass parameter, $\mu < \epsilon \, \Lambda $.
 
In the effective low-energy superpotential we get
 \beq 
 \tilde{W}= \frac{A_{3D}
M \tilde{M}}{\sqrt{2}} + \frac{ \alpha  (2 A_0^3+6 u(A_{3D}) A_0 )
}{3} - 2 \alpha \xi A_0 +2 i \alpha \mu A_0^2 \,\, ,
\label{elesu}
 \eeq
Then, the $F$ term takes the form
 \beqn
 V_F
 &=&
  e_{3D}^2\left|\frac{ M
\tilde{M}}{\sqrt{2}}+ 2 \alpha u' A_0\right|^2+ 4 e_0^2 \alpha^2 \left|  A_0^2
+  2 i \mu A_0 + u   - \xi \right|^2   
\nonumber\\[3mm]
&+&
\frac{| M A_{3D}|^2\, |\tilde{M}
A_{3D}|^2}{2} \,.
\label{efte}
\eeqn
\begin{figure}
\epsfxsize=9cm
\centerline{\epsfbox{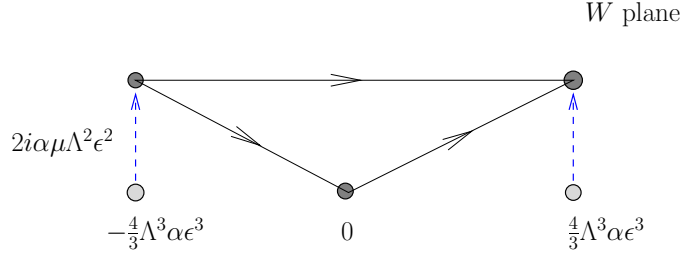}}
\caption{{\footnotesize Superpotential values in three vacua,
at nonvanishing
$\mu$ (in the  
 complex $\tilde W$ plane). }}
\label{wallssuperpotential}
\end{figure}
The Coulomb vacuum remains intact when $\mu\neq 0$
is switched on, and so is the value of the superpotential in
the Coulomb vacuum. Both confining vacua have
$A_{3D}=0$. 
The values of $A_0$ and the monopole field condensate are
 \beq 
 A_0=-\mu i  \pm
\sqrt{\xi-\Lambda^2-\mu^2}, \quad M \tilde{M}=  i 4 \sqrt{2} \alpha
\Lambda \left(-i \mu \pm \sqrt{\xi-\Lambda^2-\mu^2}\right) \,.
\label{aomm}
\eeq
The values of the superpotential  in both confining vacua change (see Fig.~\ref{wallssuperpotential}),
being shifted upwards in the complex plane,
 \beq 
 \tilde{W}_{1,2}=
\frac{2}{3} \alpha  \left[ -2 i \mu ^3 +3 i  (\xi-\Lambda^2)  \mu \mp
2 (\xi-\Lambda^2-\mu^2)^{3/2} \right].
\eeq 
The tension of the BPS domain wall is given by the absolute value of the difference of
the superpotentials at two vacua between which the given wall interpolates. 
For this reason, if the   composite BPS  walls
exist at $\mu \neq 0$, the composite wall will be stable, see Fig.~\ref{wallssuperpotential}.

\begin{figure}[h!t]
\begin{center}
$\begin{array}{c@{\hspace{.2in}}c} \epsfxsize=2.5in
\epsffile{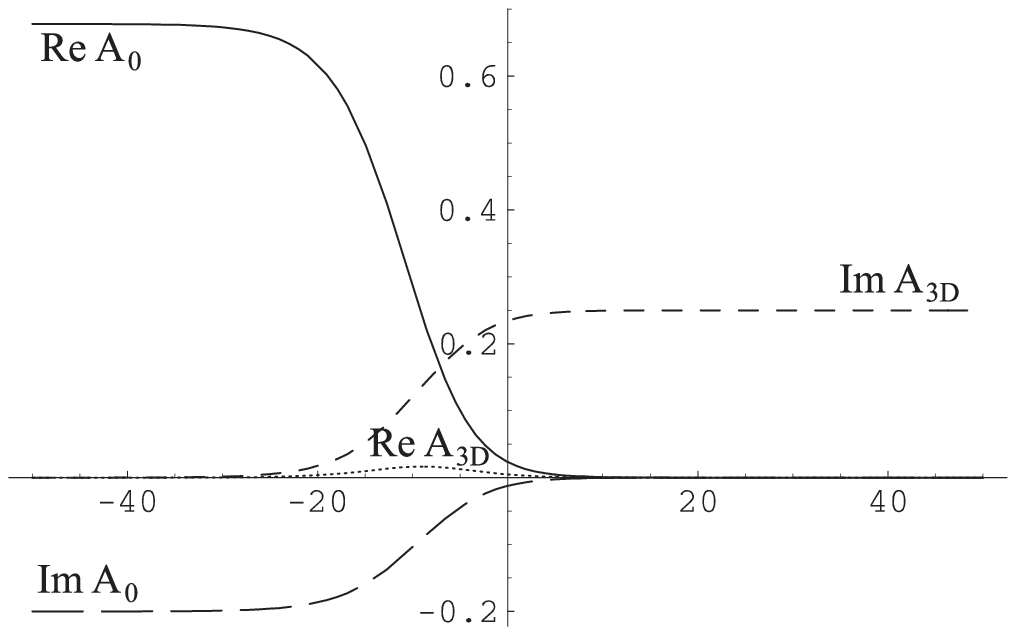} &
    \epsfxsize=2.5in
    \epsffile{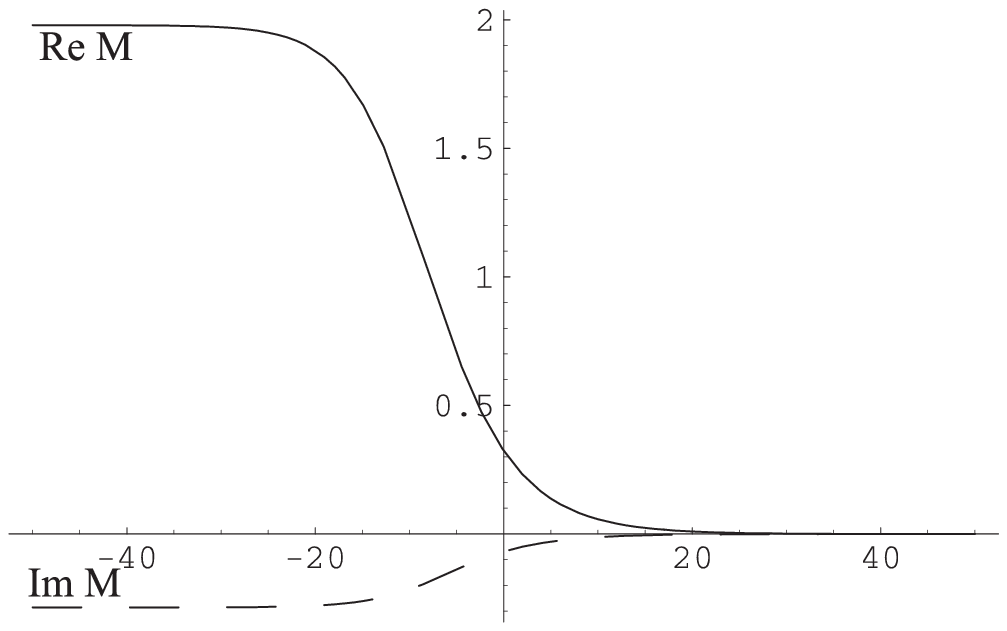}
\end{array}$
\end{center}
\caption{\footnotesize  Elementary wall profile functions at nonvanishing $\mu$.
 Left: ${\rm Re}\, A_0$ (solid), ${\rm Im} \,A_0$ (long dashes), ${\rm Im}\, A_3$ (short dashes), ${\rm Re} \, A_3$ (dots). 
 Right: ${\rm Re} \, M$ (solid) and  ${\rm Im} \, M$ (dashed).}
\label{singlemu}
\end{figure}

\begin{figure}[h!t]
\begin{center}
$\begin{array}{c@{\hspace{.2in}}c} \epsfxsize=2.5in
\epsffile{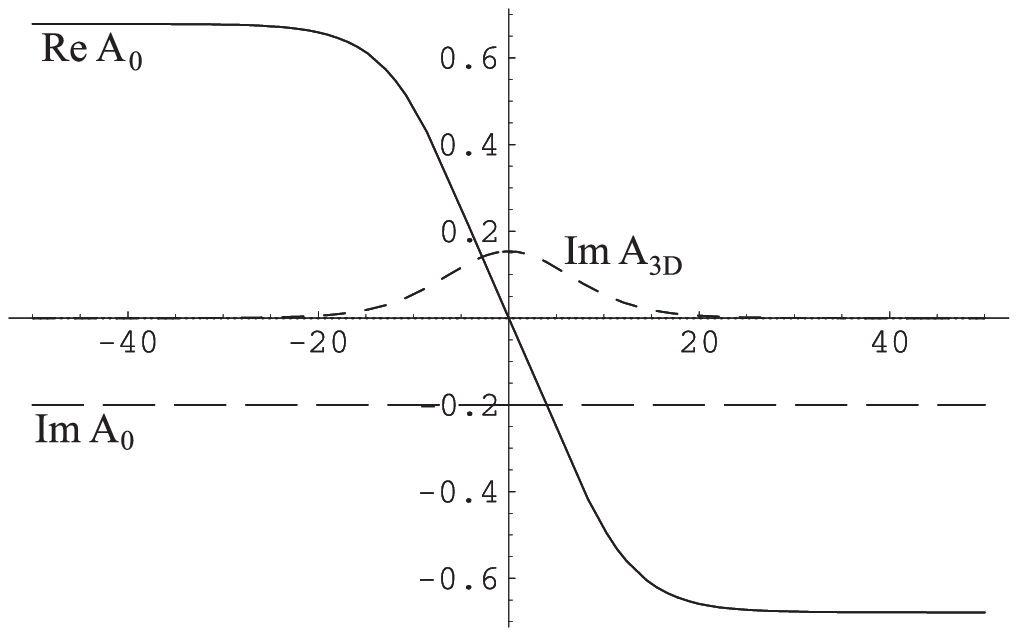} &
    \epsfxsize=2.5in
    \epsffile{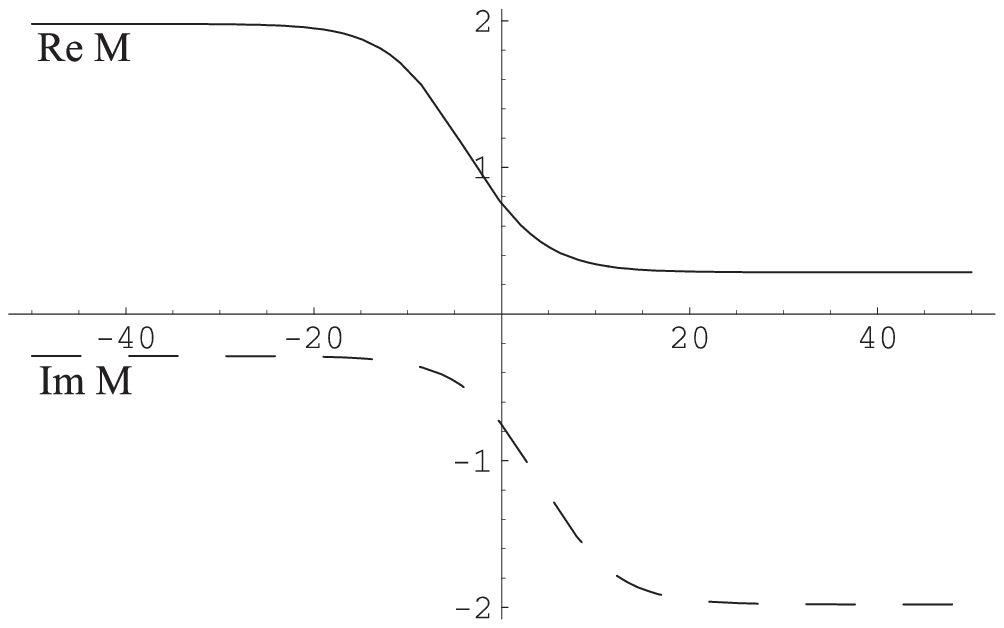}
\end{array}$
\end{center}
\caption{\footnotesize  The profile functions of the stable composite wall profiles at nonvanishing
$\mu$
 Left: ${\rm Re}\, A_0$ (solid), ${\rm Im} \,A_0$ (long dashes), ${\rm Im}\, A_3$ (short dashes).
 Right: ${\rm Re} \, M$ (solid) and  ${\rm Im} \, M$ (dashed).}
\label{doublemu}
\end{figure}

In order to write the BPS equations for the elementary wall,
we need  complex profile functions for each field, $M$, $A_{3D}$ and $A_0$.
The ansatz $\tilde{M}=i M$ can still be used.
Let us introduce a phase (see Ref.~\cite{Chibisov:1997rc} for a detailed discussion)
\beq 
\omega={\rm Arg}\, \,
 \frac{ 3 \,  (\xi-\Lambda^2) \, \mu -2  \mu ^3 }{2 \left(\xi-\Lambda^2-\mu^2\right)^{3/2}} \, . 
 \label{liap}
 \eeq
The BPS equations
 generalizing those in Eq.~(\ref{eq43}) to the case  $\mu \neq 0$ are
\beqn
&&
\frac{\partial M^\dagger}{\partial z}- i e^{i \omega} \frac{M \, A_{3D}}{\sqrt{2}} =0 \,,
\nonumber\\[2mm]
&&
\frac{\partial A_0^\dagger }{\partial z}- e^{i \omega} 2 \alpha e_0^2 (A_0^2 
 - 2  i \Lambda A_{3D} + 2  i \mu \Lambda A_{0} -\xi+\Lambda^2)=0 \, ,
\nonumber\\[4mm]
&&  \frac{\partial A_{3D}^\dagger }{\partial z}+ i e^{i \omega}  e_{3D}^2
\left( \frac{M^2}{\sqrt{2}} - 4 \alpha \Lambda A_{0}\right) =0 \, .
\label{vpfu}
\eeqn
Numerical solution for the profile functions of the elementary walls at $\mu\neq 0$
is displayed in Fig.~\ref{singlemu}.
For the composite wall we again can use the ansatz $\tilde{M}=i M$.
The BPS equations  are very similar
to the ones for the elementary walls. The only difference is
that for the composite wall $\omega=0$ and  the profiles of ${\rm Im} \, A_0$
and  ${\rm Re} \, A_{3D}$ are constant,
\beq  
{\rm Im} \, A_0=-\mu \, , \qquad  {\rm Re} \, A_{3D}=0 \, .
\label{vpfup}
\eeq
The BPS equations for the non-constant profile functions   are   
\beqn
&& 
\frac{\partial 
({\rm Re} \, M)}{\partial z}= - \frac{
({\rm Re} \,M) \, ( {\rm Im } A_{3D}) }{\sqrt{2}}  \,, 
\nonumber\\[2mm]
&&
\frac{ \partial ({\rm Im} \, M)}{\partial z}= \frac{
({\rm Im} \,M) \, ( {\rm Im } A_{3D}) }{\sqrt{2}}  \,,
\nonumber\\[2mm]
&&
 \frac{ \partial ({\rm Re} \, A_0)}{\partial z}=
2 \alpha e_0^2 (  ({\rm Re} \, A_0)^2 +  2 ({\rm Im} \, A_{3D}) \Lambda -
\epsilon^2 \Lambda^2 + \mu^2 ) \, ,  
\nonumber\\[4mm]
&&
\frac{ \partial ({\rm Im} \, A_{3D})}{\partial z}=
- e_3^2 \left(  \frac{({\rm Re} \,M)^2-({\rm Im} \,M)^2}{\sqrt{2}}
 -4 \alpha  \Lambda ({\rm Re} \, A_0)   \right) \, .
 \label{bpseqco}  
 \eeqn
From these equations we deduce  that the following quantity
remains constant ($z$-independent):
\beq 
({\rm Re} \, M )( \, {\rm Im} \, M )=-2^{3/2} \alpha \mu \Lambda 
\,.
\label{zind}
\eeq
The corresponding profiles are shown in Fig.~\ref{doublemu}.
The boundary conditions are such that $M(\infty) = -i M(-\infty)$. 
Note that if we try to use 
the boundary conditions $M(\infty) = i M(-\infty)$,
we get an unstable wall, whose tension 
is larger than twice the tension of the elementary wall.
The tensions of the BPS walls are given by the central charges,
\beqn
 T_{12}
 &=&
 T_{23}=\frac{4 \alpha}{3} \sqrt{4(\epsilon^2 \Lambda^2 -\mu^2)^3+
(3 \mu \epsilon^2 \Lambda^2 - 2 \mu^3)^2 } \, ,
\nonumber\\[3mm]
 T_{13}
 &=&
  \frac{16 \alpha}{3} (\epsilon^2 \Lambda^2 -\mu^2)^{3/2} \, , 
\label{tedw}
\eeqn
implying that the composite wall is stable.  Note also that the parameter
$\omega$ is different for the composite and elementary walls.
Two out of four supercharges will annihilate each domain wall;
but they will be different for the composite and elementary walls.

On symmetry grounds one can state that the
real and imaginary parts of the $M$ condensate  in the wall center    (i.e. at the point $z=0$) 
are equal in absolute value. Using the fact that
$({\rm Re} \, M (z))( \, {\rm Im} \, M(z) )$ is constant,
it is straightforward to analytically calculate 
the monopole condensate at the center of the composite domain wall.
 The expression for the condensate in the wall  center is very concise,
\beq 
|{\rm Re} \, M(z=0)|=|{\rm Im} \, M(z=0)| \approx 2^{3/4}
\sqrt{\mu \alpha \Lambda} \,. 
\label{veco}
\eeq
In the limit $\mu \ll \epsilon \Lambda$
the condensate outside the wall is
\beq
 |{\rm Re} \, M(z=\pm \infty)|\approx 2^{5/4} \Lambda \sqrt{\alpha \epsilon} \, ,
\quad  |{\rm Im} \, M(z=\pm \infty)| \approx 
2^{1/4} \mu \sqrt{\frac{\alpha}{\epsilon}}\,.  
\label{inwa}
\eeq
The ratio of the absolute values of the monopole
condensates inside and outside the wall is proportional to
$ \sqrt{{\mu}/{(\epsilon \Lambda )}}$.

\subsection{Confinement on the composite domain wall}
\label{cooncdw}

As in the toy model discussed in Sect.~\ref{toymodel},
we would like to understand the localization of the (massive) gauge field 
on the wall as a quasimodulus $\sigma$ localized on the wall world volume. 
Previous consideration suggests us to look for an opposite direction
 rotation of the U$(1)$ phase of the monopole field
at $z<0$ and $z>0$, respectively. 
Our target  is an exited domain wall (corresponding to $\sigma = \pi$)
in which the charged field profile vanishes in the center of the wall 
(see Fig.~\ref{stableandunstable} pertinent to the toy model of Sect.~\ref{toymodel}). 
There is an important difference between the toy model and that of Sect.~\ref{seibergwitten}. 
In the former case the Coulomb phase was 
not a true vacuum of the theory, while in the latter it is. This implies that imposing the condition
$\sigma =\pi$ we get, in fact,  two elementary walls, with no binding energy, separated by 
infinite  distance. 
Needless to say, the condition that typical energies in the low-energy theory
must be $  \ll  1/d$ cannot  be met then. 
In this formulation it makes no sense to  speak of localization
and reduction to $2+1$ dimensions.
\begin{figure}[h!tb]
\epsfxsize=10cm \centerline{\epsfbox{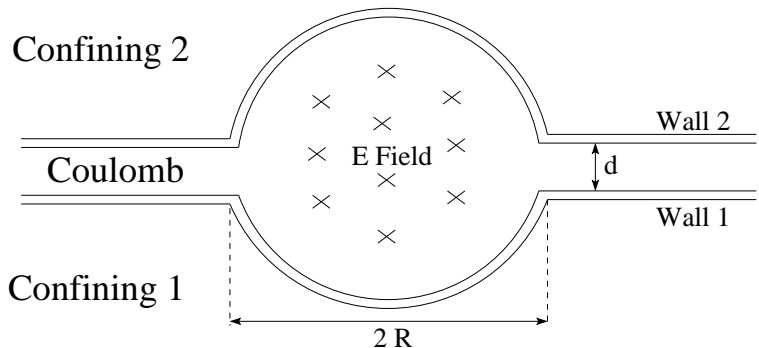}}
\caption{{\protect\small A sketch of an electric flux tube trapped 
in the middle of the composite wall. }} \label{fluxetto}
\end{figure}

To discuss confinement and localization on the composite wall we
must change the setting.
Consider the field configuration shown in Fig.~\ref{fluxetto}, which displays a
flux tube trapped in the middle of the composite wall.
Let us call $R$ the radius of the region where the electric field is localized.

The two-component wall is stabilized at a distance $d_0$
between the elementary walls. To separate 
them will cost a finite amount of energy  per unit surface of the wall, to be referred to
as $\delta T_{\rm w}$.
In the model discussed in this section, at the first nonvanishing order in $\mu$,
we have
\beq 
\delta T_{\rm w}= 6\, \alpha \,\epsilon \,\Lambda \,\mu^2 \,.
\label{delt}
 \eeq
Denote the electric flux of the vortex by $f$. An estimate for the tension of the flux tube is
\beq 
T_{\rm ft} \approx \frac{1}{2 e_{3D}^2}\,\, \frac{f^2}{\pi R^2} + 2 R \, \delta T_{\rm w} \, .
\label{eftt}
 \eeq 
Minimizing with respect to $R$ we arrive at
\beq 
R_* =\left( \frac{f^2}{2 \pi \, e_{3D}^2 \, \delta T_w } \right)^{1/3} \, .
\label{mwrr}
\eeq
This shows that $R_*$ is a finite quantity, and the flux is indeed squeezed into flux tubes
inside the composite domain wall.
This is, of course, applicable provided that $\delta T_{\rm w}$ is   positive 
(i.e. the force between two elementary domain walls is attractive, which is the case
in the model at hand). 

Now, let us consider the case of a constant electric flux
per unit  length of the wall in the perpendicular direction
(analogous to the picture in Fig.~\ref{constantb})
We will denote this quantity --- flux
per unit  length --- by the same letter
 $f$.
This flux is trapped in the middle Coulomb vacuum.
As a consequence, there is a repulsive contribution to the potential
between the two components of the composite wall,
\beq 
T_{\rm wf} = \frac{1}{e_{3D}^2 }\,\, \frac{f^2}{d}\,. 
\label{potef}
\eeq
In order to find the minimum of the overall potential, we have to add this
to the domain wall tension $T_{\rm w}$ considered as a function of $d$. 
At large distance $T_{\rm w}$ has just an exponential tail in $d$; therefore, 
the-long range force is repulsive, as follows from Eq.~(\ref{potef}).
If the flux $f$ is small, however,  the global minimum of the
inter-wall potential is still at finite $d$.
There is a critical value of $f$ --- let us call it $f_0$ --- for which the elementary domain walls
at the equilibrium 
will be separated by a infinite distance. 
A simple energy estimate can be used to evaluate the 
value of $f_0$,
\beq \frac{1}{e_{3D}^2 } \,\,\frac{f^2_0}{d_0} > \delta T_{\rm w} \, , \quad
 f_0 = e_{3D} \sqrt{d_0  \, \delta T_{\rm w}}\,,
\label{simes}
\eeq
where $d_0$ is the domain wall separation at zero $f$.

Thus, although the low-energy description in the Seiberg--Witten motivated model
at hand  is
not of the sine-Gordon type (cf. (\ref{effectiveaction})), 
the quasimodulus-based low-energy description is still
valid at $|\sigma |\ll \pi$: a mass term $m\sigma^2$ is generated.

\section{ Two-Flavor Model}
\label{2flavors}

\subsection{$\mathcal{N}=2$ SQED with two flavors}
\label{n2sqed2}

At this point one is tempted to say that the fact
that the gauge field 
localized on the wall weakly confines is due to residual exponentially small
VEV's of the Higgs field inside the wall. In this case, the 
absence of the Coulomb regime on the wall would be a universal phenomenon. 
In fact, we know that this is not the case \cite{Shifman:2002jm} (see also the review paper
\cite{Shifman:2007ce}). In Ref.~\cite{Shifman:2002jm}
$\mathcal{N}=2$ SQED with two flavors was considered.
This model has a domain wall with a phase field $\sigma$
localized on it. This field is a Goldstone of a spontaneously broken
U(1) and, thus, remains exactly massless.
Upon Polyakov's dualization it represents a U(1) gauge field on the wall in the 
Coulomb regime. Moreover, in the same paper \cite{Shifman:2002jm}
it was shown that a magnetic flux tube
coming from the bulk in the perpendicular to the wall direction ends on the
wall creating a vortex of the $\sigma$ field. Upon Polyakov's dualization this vortex
is interpreted as an electric charge, a source of the electric field
on the wall. Two such sources interact through the Coulomb potential 
at large distances (logarithmic in 2+1 dimensions).

Residual exponentially small
VEV's of the Higgs field inside the wall are certainly important.
For instance, in the toy model considered
in Sect.~\ref{toymodel} the $\sigma$ field turns out to be a quasimodulus
only because the wall at hand is very thick implying exponential suppression of the 
Higgs field inside the wall. If the wall was thin, $\sigma$ would be heavy, and the only
light field on its world volume would be the translational modulus.
In the two-flavor model
we can introduce $\sigma$ as a relative phase between
$q_1$ on one side of the wall and $q_2$ on the other side.
Because these fields are different, there is no need in a $z$ dependent interpolation of $\sigma$
inside the wall, as was the case in the one-flavor model. This is the technical reason for its masslessness.

Below we will illustrate the emergence of the {\em exact} moduli field $\sigma$
in $\mathcal{N}=2$ SQED with $2$ flavors analyzing it in a regime somewhat 
different from that of Ref.~\cite{Shifman:2002jm}. This moduli field remains
massless even in the limit of thin wall.
Thus, the mode of implementation
of 2+1 dimensional electrodynamics on the wall world volume
--- Coulomb vs. confinement --- is a dynamical issue. 

Consider $\mathcal{N}=2$ SQED with two
flavors and the matter mass terms chosen as follows:
\beq
m_1 =m\, , \qquad m_2 =-m\,, \quad \Delta m \equiv m_1-m_2 \equiv 2m\,.
\label{mater}
\eeq
We will introduce the Fayet--Iliopoulos term $\xi$ through
the superpotential. Then
the bosonic part of the action can be written as
\beq
 S=\int d^4 x \left\{ \frac{1}{4 e^2} F_{\mu \nu}^2 +
\frac{1}{e^2} |\partial_\mu a|^2 +\bar{\nabla}_\mu \bar{q}_A \nabla_\mu q^A +
\bar{\nabla}_\mu \tilde{q}_A \nabla_\mu \bar{\tilde{q}}^A
 +
V_D+ V_F \right\},
\nonumber\\[3mm]
\mbox{
}
\label{n2sqed}
\eeq
where the scalar potential is the sum of $D$ and $F$ terms,
\beqn
 V_D
 &=&
 \frac{e^2}{8}\left(|q^B|^2-|\tilde{q}_B|^2\right)^2\,,
\nonumber\\[2mm]
 V_F
 &=&
 \frac{1}{2} \left |q^B(a+\sqrt{2}m_B)\right|^2+
\frac{1}{2}\, \left|\tilde{q}_B ( a+\sqrt{2}m_B)\right|^2
\nonumber\\[2mm]
&+&
\frac{e^2}{2}\left|\tilde{q}_A q^A- \frac{\xi}{2}\right|^2\,.
\label{potgauge}
\eeqn
This theory has two vacua, and a domain wall interpolating between them.
It was fully analyzed in \cite{Shifman:2002jm} in the limit of
thick wall
\beq
m\gg e\sqrt\xi\,,
\label{largem}
\eeq
when the overlap between two edges of the wall is exponentially small.

Now we will discuss the same problem in the opposite limit
of a thin wall, with a strong overlap,
\beq
m\ll e \sqrt{\xi}\,.
\eeq
This limit is usually referred to as the sigma-model limit.
If  
$ m\ll e\sqrt\xi$,
the ``photonic" supermultiplet becomes heavy,
since the photon mass in the bulk $\sim e\sqrt\xi$.
Therefore, it can be integrated out, leaving us with
the theory of  fields from the matter supermultiplets, nearly massless in the scale $ e\sqrt\xi$,
which interact through a nonlinear sigma model with the K\"ahler term
corresponding to the Eguchi--Hanson metric. 
The manifold parametrized by these (nearly) massless fields
is  four-dimensional. The two vacua of the model  vacua  lie at the base of this
manifold. Therefore, in considering the domain wall solutions
in the sigma-model limit $ m\to 0$ \cite{GTT,Tw,GPTT} 
one can use the ansatz $q=\tilde{q}^\dagger$
and limit oneself to the base manifold\index{base manifold}, 
which is, in fact, a two-dimensional sphere.
In this way we arrive at the problem of the domain wall in the CP(1) model
deformed by a twisted mass term (related to a nonvanishing $\Delta m =2m$).
In this formulation the problem was first addressed in \cite{Tw}.

In the sigma-model limit one can readily find
explicitly the wall profiles,
\beqn
q^1
&=&
\bar{\tilde{q}}^1=\sqrt{\frac{\xi}{2}} \left(\cos
\frac{\eta(z)}{2}\right)\,,
\nonumber\\[3mm]
q^2 &=&
\bar{
\tilde{q}}^2=\sqrt{\frac{\xi}{2}} \left(\sin \frac{\eta(z)}{2}\right)
e^{i \sigma}\,, \nonumber\\[3mm] a &=& m \sqrt{2}  \left(\sin^2
 \frac{\eta}{2} - \cos^2 \frac{\eta}{2}\right)= -m \sqrt{2} \, \cos
\eta\, ,  \label{eq33}
\eeqn
where
\beq
\eta(z)=2 \arctan\, (\exp(2 m z))\,.
 \label{etta}
  \eeq
Note that 
\beq |q^1|^2 + |q^2|^2= {\xi/2}
 \eeq 
for all $z$.
The modulus $\sigma$ in Eq.~(\ref{eq33}) reflects the fact that the target space
of the CP(1) model with the twisted mass has U(1) symmetry. It is spontaneously
broken on each given wall solution. More details on kinks in the 
CP(1) model with the twisted mass,
which appear as domain walls in the problem at hand,
can be found in \cite{Shifman:2007ce}.

Since $e^2\to\infty$  in the sigma-model limit, in the bulk action we can neglect
 the contribution due to the gauge field strength tensor.
The gauge field then becomes non dynamical; it is expressible 
in terms of the matter fields,
\beq
 A_\mu=\frac{i \left(\bar{q}  \stackrel{\leftrightarrow}{\partial_\mu} q  -
\bar{\tilde{q}} \stackrel{\leftrightarrow}{\partial_\mu} \tilde{q}\right)}
{\bar{q} q+\bar{\tilde{q}} \tilde  q}\,.
 \label{tizio}
\eeq
The scalar field 
kinetic term tends to zero too, implying
\beq
a=\sqrt{2} \,
m \,\,
\frac{|q_2|^2+|\tilde{q}_2|^2-|q_1|^2-|\tilde{q}_1|^2 }{\bar{q} q+\bar{\tilde{q}} \tilde  q} \, . \label{caio}
\eeq

As usual, we promote $\sigma$ to a $(x,y,t)$-dependent field on the wall world-volume.
In our gauge the field $A_z$ vanishes while the nonvanishing component are 
\beq 
A_k = - 2  \sin^2 \frac{\eta(z)}{2}   (\partial_k \sigma) \,,\qquad k=1,2\, .
\label{novc}
\eeq
The field strength then takes the form
\beq 
F_{ k z}= \frac{2 m}{\cosh^2 2 m z} \, (\partial_k \sigma) \,,\qquad k=1,2\, . 
\label{resi} 
\eeq
If we  substitute these expressions back  in the action, we get
\beq
S=\int dz\, dt\, d^2 x
 (\partial_k \sigma)^2   \left(
\frac{\xi}{2 \cosh^2 2 m z} +   \frac{2 m^2}{e^2  \, \cosh^4 2 m z} \right) \, , 
\eeq
the the first term is due to the covariant derivative 
and the second term is due to the field strength tensor.
The second term is negligible and can be omitted.
This  shows that our approximation is self-consistent.
Keeping only the first term and integrating over $z$
gives us the normalization of the world-volume effective action,
\beqn
S&=&\int d^3 x \, \beta \, (\partial_k \sigma)^2 \, , \nonumber\\[3mm]
 \beta&=&\int dz \, \frac{\xi}{ 2 \cosh^2 2 m z} = \frac{\xi}{2 m} \, . 
 \label{nwvea}
 \eeqn
 The $\sigma$ field  is strictly massless,  
the gauge field that dualizes $\sigma$ 
is in the Coulomb phase on the wall world volume.

Quantitatively, the thin wall (sigma model) approximation
is not parametrically supported. However,
it reveals the existence of a massless modulus on the wall in the most straightforward way.
Moreover, it can be conveniently used to discuss a ``boojum" configuration, with a flux tube ending on
the wall. The approximation is analytic at sufficiently large
distances from the point of the wall-tube junction.
 
The wall-tube junction solution in the sigma-model limit
was found (in a different notation) in Refs.~\cite{GPTT,INOS}.
For illustrative purposes we will reproduce it in our notation.
The BPS equations for a vortex ending on the wall  are (see Ref.~\cite{Shifman:2002jm})
\beqn
&&
B_3-\frac{g^2}{2} \left( 2 |q_k|^2-\xi\right) - \sqrt{2} \partial_3 a =0 \, , \quad
 B_1-i B_2 -\sqrt{2} (\partial_1-i \partial_2) a = 0 \, , \nonumber\\[2mm]
&&
 \nabla_3 q_k=- \frac{1}{\sqrt{2}}q_k (a+\sqrt{2} m_k) \, , \quad
(\nabla_1-i\nabla_2) q_k=0 \, .
\label{eqboojum}
\eeqn
In the sigma-model limit only $q_k$'s are dynamical variables. Thus,
 we need to solve two equations in the second line in Eq.~(\ref{eqboojum}).
This can be done using the following ansatz in 
the cylindrical coordinates $z,r,\phi$:
\beq 
q_1=\sqrt{\frac{\xi}{2}} \, e^{i \phi} \, \cos \frac{\theta(z,r)}{2} \, ,
 \qquad q_2 = \sqrt{\frac{\xi}{2}}  \, \sin \frac{\theta(z,r)}{2} \, .  
 \label{cyl} 
  \eeq
Then the BPS equations can be written as
\beq 
\frac{\partial \theta}{\partial r}= \frac{1}{r} \sin \theta \, , \qquad
\frac{\partial \theta}{\partial z}= 2 m\,  \sin \theta \, .
\label{cylp}
\eeq
The solution can be readily found, namely,
\beq 
 \theta= 2 \arctan  \, \left[ \frac{ \, \exp \left(2 m (z-z_0)\right)}{ r} \right] \, ,
 \label{cylpp}
 \eeq
where $z_0$ is an integration constant which parametrizes the $z$ position of the object.
Now
 we can use Eqs.~(\ref{tizio}) and (\ref{caio}) to determine $A_\mu$ and $a$,
\beq a
=-\sqrt{2} m  \cos \theta \, , \qquad A_{\hat{\phi}}=-\frac{\cos^2 (\theta/2)}{r} \, .
\eeq
This solution is valid in the sigma-model limit ($m\ll e \sqrt{\xi}$) and 
at distances $r\gg {1}/({e \sqrt{\xi}})$  from the point of the wall-tube junction. We see that the gauge field is localized inside the wall and has the Coulomb $1/r$
behavior at large $r$. This shows that we do have the Coulomb phase
on the wall -- even in the sigma model limit -- and confirms that $\sigma$ is 
a strictly massless modulus.

To conclude this section let us return to the gauge theory limit (\ref{largem}) of the
problem at hand studied in \cite{Shifman:2002jm}.
One might naively suspect that there are two different moduli fields in the two-flavor model.
One is the  modulus $\sigma$ which is exactly massless because it is related to the global
U(1) symmetry broken by the wall solution. One might guess that this modulus has nothing
to do with the bulk gauge field. Allegedly, the bulk gauge field localized on the wall
through the same mechanism as was discussed in the one-flavor model is a quasimodulus 
$\tilde{\sigma}$ which is different from the modulus $\sigma$, and 
$\tilde{\sigma}$ acquires a small mass due to  exponentially small
quark fields inside the wall, as in Sect. \ref{toymodel}. 

The solution for the wall-string junction found
in \cite{Shifman:2002jm} shows that this naive picture is incorrect. In \cite{Shifman:2002jm}  
it was shown
that the string orthogonal to the domain wall can end on the wall, and the
magnetic flux it carries penetrates in the wall. The endpoint of the string
plays the role of a vortex for the $\sigma$ field localized on the wall. In fact, 
the solution for the wall (far away from the string endpoint) 
is approximately given by the 
unperturbed domain wall solution with the collective coordinate $\sigma$
determined by  \cite{Shifman:2002jm}
\beq
\sigma \,=\, \alpha,
\eeq
where $\alpha$ is the polar angle on the wall plane. 
The endpoint of the string creates a vortex of the field $\sigma$. This shows that
in fact there is no extra field $\tilde\sigma$.
The gauge field localized on the wall is dual to 
$\sigma$ which is strictly massless in the two-flavor model.

\subsection{Further discussion of moduli vs. quasimoduli}

The example of Sect.~\ref{n2sqed2} shows  that the two-flavor
model has a massless phase field $\sigma$ localized on the domain wall.
This is not related to supersymmetry. Any model with a global U(1) symmetry, two 
distinct vacua in which the global symmetry is unbroken, and a domain wall that
spontaneously  breaks this symmetry, automatically has  a massless Goldstone boson 
localized on the domain wall. 
In particular, the wall solution with $\sigma =\pi$
has the same tension as that with $\sigma = 0$, in an obvious
contradistinction with the solution of the toy model
discussed in Sect.~\ref{toymodel}. Note that in the latter case, in the 
$\sigma =\pi$ wall the matter field vanishes on the plane lying in the middle of the wall.
No such zero-plane occurs in the former case.

To better understand the relation
between the one- and two-flavor models, we should dwell on the following question:
why there is no confinement
on the wall in the two-flavor models, although the matter field
condensates do not exactly vanish inside the wall?
Why the $\sigma$ field remains massless in this case?
A (partial) answer to  this puzzling question  is as follows.
3+1 dimensional physics inside the wall, between its edges, is
not the only thing to consider. Existence vs. nonexistence
of a massless modulus is a global effect.
The boundary conditions at $z=\pm \infty$ play a crucial role.
In the two-flavor model we deal with four phases:
\beq
q_1(z=-\infty), \,\,\, q_2(z=-\infty)  \,\,\, {\rm and} \,\,\,  q_1(z=+\infty), 
\,\,\, q_2(z=+\infty)\,.
\eeq
Moreover, 
\beq
q_1(z=-\infty) \to \sqrt{\frac{\xi}{2}}\,,\quad {\rm and}\,\,\, 
 q_2(z=+\infty) \to \sqrt\frac{\xi}{2}\,,
\eeq
 while 
 \beq
| q_2(z=-\infty)|\to  0\,,\quad
|q_1(z=+\infty)| \to 0\,. 
\eeq
The massless modulus $\sigma$ corresponds to a rotation of
$q_1(z=-\infty)$ and $q_1(z=+\infty)$ in the same direction while $q_2(z=-\infty)$ and
$q_2(z=+\infty)$ remain fixed.  
 
The would-be quasimodulus of Dvali {\em et al.} corresponds to a rotation of the phases
of $q_1(z=+\infty)$ and  $q_2(z=+\infty)$ in the same direction while $q_1(z=-\infty)$
and $q_2(z=-\infty)$ remain fixed. The would-be quasimodulus has a tachyonic direction and classically
decays into the modulus $\sigma$.

\begin{figure}[h!tb]
\epsfxsize=14cm \centerline{\epsfbox{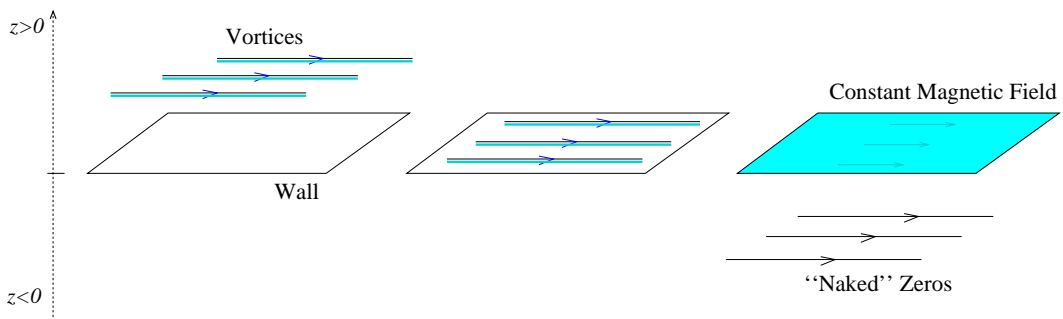}}
\caption{{\protect\small A grid of flux tubes parallel to the domain wall. 
At $z>0$ they are made out of the $q_2$ field condensate. 
The line of zeroes of the $q_2$ field is surrounded by the 
magnetic field. At $z=0$ the vortices lies in the middle of the wall. 
Proceeding further the zeroes of $q_2$ go to $z$ negative while 
the magnetic field remains on the domain wall surface. 
In the limit where the zeroes are at $- \infty$ we recover the solution 
describing a constant magnetic field on the wall.}} 
\label{parallel}
\end{figure}

Let us try to understand  mechanisms
responsible for this phenomenon in more detail.
Consider a string parallel to the domain wall (a grid of such strings is depicted
in Fig.~\ref{parallel}).
If the distance between the wall and the string is very large we have an ordinary ANO 
string with thickness $\sim {1}/{(e\sqrt{\xi})}$. As we move 
the string toward the wall, the quark condensate decreases and, 
in the center of the wall, the thickness of the string
(in the directions parallel to the wall) becomes 
\beq 
\sim R = \frac{1}{e \sqrt{\xi}e^{-dm/2}} \,,
\label{tsdpw}
\eeq
where $d$ is the thickness of the wall. 
The thickness of the wall depends on the regime 
in which we find ourselves. 
In the limit ${m}/{(e\sqrt{\xi})} \gg 1$ the thickness 
is entirely determined by the matter field  and is $d\sim {m}/{(e^2 \xi)}$. In the 
opposite limit ${m}/{(e\sqrt{\xi})} \ll 1$ (the sigma-model limit) the thickness is $d \sim {1}/{m}$. 

\begin{figure}[h!tb]
\epsfxsize=8cm \centerline{\epsfbox{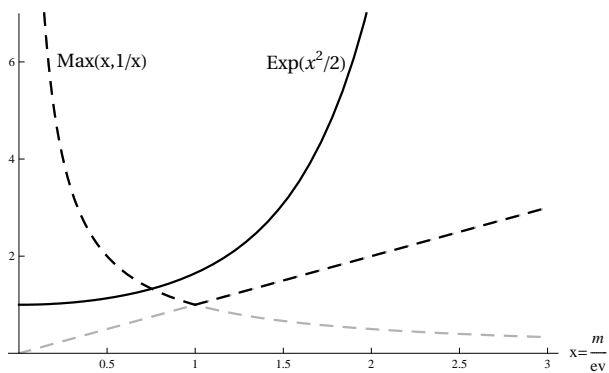}}
\caption{{\protect\small Comparison between the thickness of the domain wall, and the transverse
size of the flux tube located inside the domain wall. In the region $x \ll 1$ corresponding to the 
sigma-model limit, the transverse
size of the flux tube is much smaller than the wall thickness.}} \label{graf}
\end{figure}

If we want to compare the thickness of the domain wall with that of the string 
in the middle of the wall we should compare 
$$
{\rm Max}\left[\frac{1}{m},\frac{m}{e^2 \xi}\right]
$$ 
with $$\frac{1}{e \sqrt{\xi}e^{-dm/2}}\,.$$ 
Multiplying both sides by $e\sqrt\xi$ we can express everything in terms of 
a dimensionless parameter 
 \beq
 x = \frac{m}{e \sqrt{\xi}}\,.
 \label{dipa}
 \eeq
Thus, we should  compare 
$${\rm Max}\left[\frac{1}{x},x\right]\,\,\,{\rm with}\,\,\,
e^{\frac{x^2}{2}}\,.$$
 These two functions are plotted in Fig.~\ref{graf}.

In the $x\gg 1$ limit the thickness of the string is much larger than the 
thickness of the domain wall. This is qualitatively consistent with the existence of a Coulomb phase in the middle of the wall. The opposite limit $x \ll 1$ is problematic. The thickness of the string is much smaller than the thickness of the wall. We are thus tempted to conclude that in this regime the wall lives in a confining phase, which is clearly in contradiction with the rigorous proof  above of the existence of a massless modulus and, hence, the Coulomb phase.

The way out of this paradox is as follows.  In the previous discussion we have compared the thickness of the wall with that of the string in the {\it middle} of the wall. So we (erroneously)  assumed that 
dynamics on the wall world volume is directly deducible  from consideration of 3+1
dimensional dynamics  inside the wall, between its
two edges.  In the $x \gg 0$ limit this creates little problem. In this limit we have a thick region
inside the wall in which the matter field condensate  essentially vanishes.  But in the $x \ll 1$ 
 limit this  way of thinking creates a paradox and, thus, reveal its inconsistency. 

In fact, the wall world-volume dynamics reflects not only what happens
at $z \sim 0$, but also what happens at  $z \to \pm \infty$.   

Consider a grid of flux tubes at $z >0$ parallel to the domain wall
(Fig.~\ref{parallel}). We should remember that $z>0$ is 
the region where $q_2$ condenses while $z<0$ the region 
where $q_1$ condenses.  Each flux tube has a line of zeroes for 
the $q_2$ field and a magnetic field surrounding this line of zeroes.  
We then move the grid of flux tubes toward the domain wall.  We want to 
understand what happens to these  flux tubes as we pass through 
the wall and then move toward negative infinity. At $z=0$ the flux 
tube grid  is in the middle of the wall; the thickness of each flux tube is larger. 
If we move on, something new 
happens. The lines of zeroes go to $z$ negative and get separated from 
their magnetic field. The magnetic field remains trapped inside the wall. In the 
limit where the zeroes are at $z \to -\infty$, we 
recover the solution of constant magnetic field inside the domain wall. 
So, the $q_2$ flux tube, passing through the wall to the other side of 
the wall, {\it does not} become the $q_1$ flux tube.  

This is not in contradiction with the symmetries of the theory. 
The domain wall is symmetric under the $Z_2$ transformation 
$z \leftrightarrow -z$ combined with $q_2 \leftrightarrow q_1$. 
The one-flavor case is different. The domain wall is symmetric under 
the parity transformation $z \leftrightarrow -z$ and the $q$ flux 
tube passing through the wall to the other side of the wall preserves its 
``identity"  remaining the $q$ flux tube.

Of course, physically the line of the $q_2$ zeroes at 
$z=-\infty$ (more exactly, the plane of the $q_2$ zeroes~\footnote{ A flux tube
has line of zeroes of a charged scalar inside it, while in order to have
a Coulomb phase on the wall we need the whole plane of the $q$
zeroes (parallel
to $x,y$-plane).}) is in fact not so ``far away."
The $q_2$ quark has the exponential profile 
$\sim \exp{(-e^2\xi\,z^2)}$ inside the wall and 
essentially becomes zero at distances of the order of $1/e\sqrt{\xi}$. 
Thus, the plane of the $q_2$ zeroes is shifted from
the region where the magnetic field is concentrated by  separation
of the order of thickness of the flux tube in the vacuum. This is in
accord with our physical intuition.

Here we arrive at a crucial distinction of the two-flavor model
from the one-flavor model of Sect.~2. In the one-flavor model
the ``empty" domain wall (i.e. without magnetic field) has a nonvanishing $Q$
field everywhere. Inside the wall it becomes small,
but still does not vanish. In 
order for a magnetic field to penetrate the wall we need to have
zeroes of the $Q$-field. Clearly, it costs less energy to create
a line of $Q$-zeroes in the $(x,y)$-plane than the whole $(x,y)$-plane 
of the $Q$-zeroes. This qualitatively explains why
we have confinement on the wall in the one-flavor model ($\sigma$
is a {\em quasi}modulus) and the Coulomb phase on the wall in the 
two-flavor model ($\sigma$ is strictly massless).

Now, let us discuss energetics of this process. First of all, consider
the string grid  when it is  far away from the wall in the 
positive-$z$ region. Assume we deal
with a homogeneous grid with density $f$, so that the flux per 
unit of length is $4 \pi f$. The tension of this configuration is the sum 
of the wall tension plus the string tensions
\beq
T_{z \gg 0} =  \xi \Delta m + 2 \pi \xi f \,.
\eeq
In the opposite position, when the zeroes are at negative $z $, far away from the wall, 
we can also easily 
compute the tension. It is just that of a domain wall with a constant magnetic flux inside it. 
An easy way to get the result is the thin-edge approximation 
(similar to that adopted in \cite{Bolognesi:2007zz} for the $Q$-wall). 
The wall tension is now given by a sum of three terms,
\begin{equation}
T\left( d \right) =\frac{1}{d} \left( \frac{2 (\Delta m)^{2} }{g^{2} } 
+ \frac{8 \pi^2 f^2}{g } \right)
+\frac{g^{2}\xi^{2}}{8} d \,.  
\label{3terms}
\end{equation}
Minimizing with respect to $d$ we obtain
\begin{equation}
d = \frac{4 }{g^2\,\xi } \left( (\Delta m)^2 + 4 \pi^2 f^2 \right)^{1/2} ,  
\label{segt}
\end{equation}
and
\begin{equation}
T_{z \ll 0} = \xi \sqrt{ (\Delta m)^2 + 4 \pi^2 f^2 } \,.  
\label{segtp}
\end{equation}

\vspace{3mm}

\begin{figure}[h!t]
\begin{center}
$\begin{array}{c@{\hspace{.2in}}c} \epsfxsize=2.5in
\epsffile{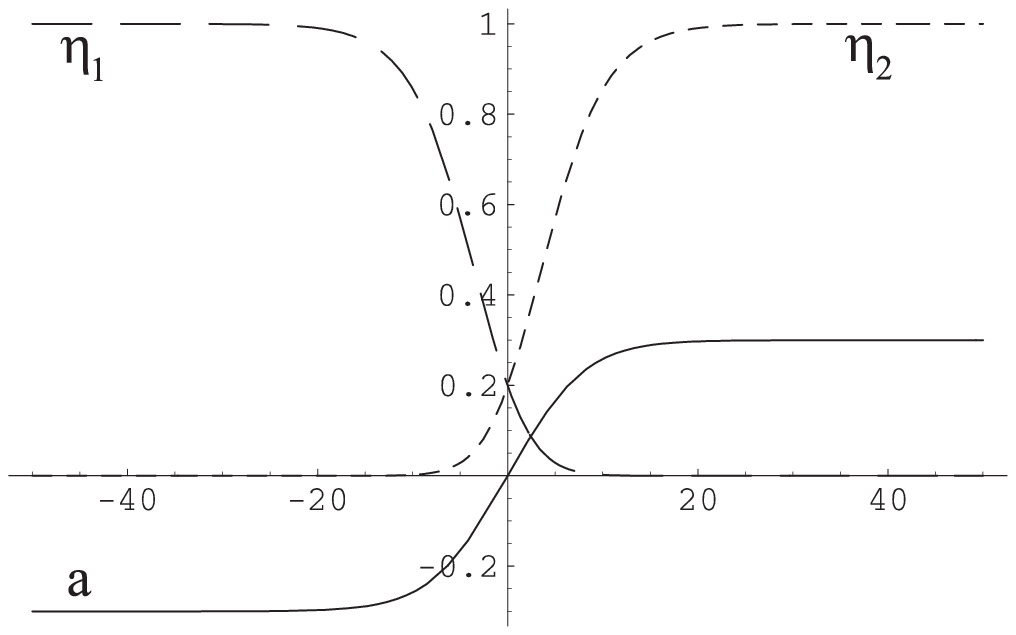} &
    \epsfxsize=2.5in
    \epsffile{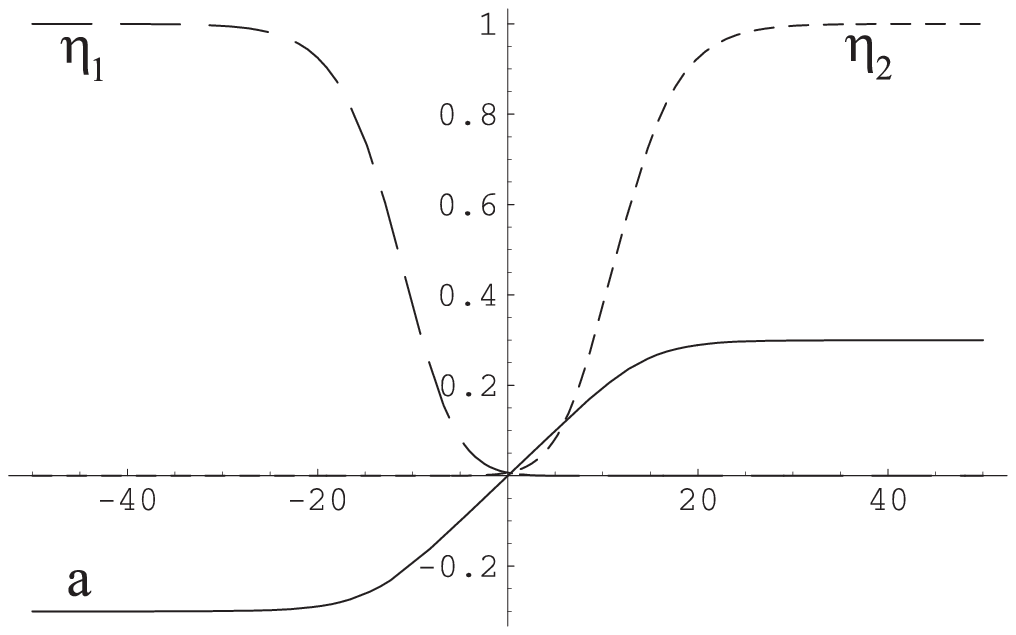}
\end{array}$
\end{center}
\caption{\footnotesize Left: Domain wall without magnetic flux ($\alpha=0$).
  Right: Domain wall with magnetic flux ($\alpha=\pi/3$). Due to the magnetic
  field, the domain wall thickness increases. The following numerical values were used:
  $m=0.3$, $e=0.2$, $\xi=1$.}
\label{magnetic}
\end{figure}

The expression in Eq.~(\ref{segtp})
can also be obtained from a more rigorous derivation using the  Bogomol'nyi 
completion method (see Ref.~\cite{Bo}).
The physical situation is very similar to the $Q$-kinks 
discussed in detail in Refs.~\cite{Abraham:1992vb} for $1+1$ dimensional sigma models.
Following \cite{Shifman:2002jm} we denote 
\beq
\label{qqt}
q^A=\bar{\qt}_A\equiv \frac{1}{\sqrt{2}}\,\vp^A\, ,
\eeq
where we introduce a new complex field  $\vp^A$.
The action then reduces to 
\begin{eqnarray}
S_{\rm red } &=&\int d^4 x \left\{ \frac1{4\gs}F_{\mu\nu}^2 
+\frac1{\gs}|\partial_{\mu}a|^2
+\bar{\nabla}_{\mu}\bar{\vp}_A\nabla_{\mu}\vp^A
\right.
\nonumber\\[3mm]
&+&\left. \frac{g^2}{8}\left( |\vp^A|^2 -\xi \right)^2
+\frac12 \left|\vp^A\right|^2\, \left| a+\sqrt{2}m_A\right|^2
\right\}\,.
\label{redqed}
\end{eqnarray} 
The Bogomol'nyi completion of the wall$+$tube energy functional can be written as
\begin{eqnarray}
T_{\rm w} = \int dz  
&& \left\{ \,\, 
\left|\ca \nabla_z \vp^A\pm  \frac1{\sqrt{2}}\vp^A(a+\sqrt{2}m_A)\right|^2 \right. \nonumber \\
&&+ \left| \sa \nabla_z \vp^A \mp i \, \nabla_x \vp^A  \right|^2 \nonumber\\
&&+ \left|\frac1{g}\pz a \pm \ca \frac{g}{2\sqrt{2}}(|\vp^A|^2-\xi)\right|^2  \nonumber \\ 
&& +\left|\frac1{\sqrt{2} g} \partial_z A_x \pm \sa \frac{g}{2\sqrt{2}}(|\vp^A|^2-\xi)\right|^2  \nonumber \\
&&\pm  \left. \frac1{\sqrt{2}} \ca   \xi\pz a  \pm  \frac1{2} \sa   \xi \pz A_{x}  \right\}.
\label{bog}
\end{eqnarray}
The BPS equations are obtained by putting to zero
each of the first four lines of Eq.~(\ref{bog}). 
In order to find an explicit solution, 
let us choose a gauge where $A_z=0$; the following ansatz can then be used:
\beq \varphi_k=\eta_k(z) \exp (i \lambda_k x) \, , \qquad  a=a(z) \, ,  \qquad A_x=-f(z) \, .\eeq
From a particular linear combination of the BPS equations we can find the value of $\lambda_k$
and also that the profile for $a$ and for $f$ are proportional:
\beq f(z)=\sqrt{2} \tan \alpha \, a(z) \, , \qquad \lambda_k = m_k \tan \alpha \, .\eeq
The equations for the other profiles give the following first order system:
\beq \sqrt{2} \, \partial_z a + \frac{g^2}{2} \cos \alpha (\eta_1^2+\eta_2^2 - \xi) = 0 \, , \eeq
\[ \partial_z \eta_k+ \frac{1}{\sqrt{2} \cos \alpha} \eta_k (a+\sqrt{2} m_k)=0 \, . \]
A numerical solution is shown in Fig.~\ref{magnetic}.

For the moment $\alpha$ is  an arbitrary angle (this is a usual trick used in analyzing dyons or 
$Q$-kinks). The wall$+$tube boundary conditions are as follows. At $z \to +\infty$
\beq 
\vp^1=0  \ , \quad \vp^2=\sqrt{\xi}  e^{i 2\pi f} \ , \quad A_x = 4 \pi f\,.
\eeq
At $z \to -\infty$
\beq 
\vp^1=\sqrt{\xi}  \ , \quad \vp^2=0  \ , \quad A_x = 0\,.
\eeq
The decomposition (\ref{bog}) give   the upper bound
\beq
\label{bound}
T_{\rm w} \geq \ca \xi \Delta m + \sa 2 \pi \xi f \,.
\eeq
Maximizing with respect to $\alpha$ we get exactly the expression in Eq.~(\ref{segtp}).]

This presents a more quantitative proof of our assertion. The $q_2$ string parallel to the wall reaches the minimum of the energy when the line of zeroes of $q_2$  is at $ z \to -\infty$, and the tension is given by (\ref{segtp}).  If, instead, we consider a set of $q_1$ strings parallel to the wall, the minimum is reached when the line of zeroes of $q_1$ is at $z \to + \infty$. In both cases the energy is always localized inside the domain wall in the form of  a constant magnetic field. The constant magnetic field corresponds to the Coulomb phase on the wall world volume, with the massless modulus  $\sigma = k y$. 

\begin{figure}[h!b]
\epsfxsize=12cm \centerline{\epsfbox{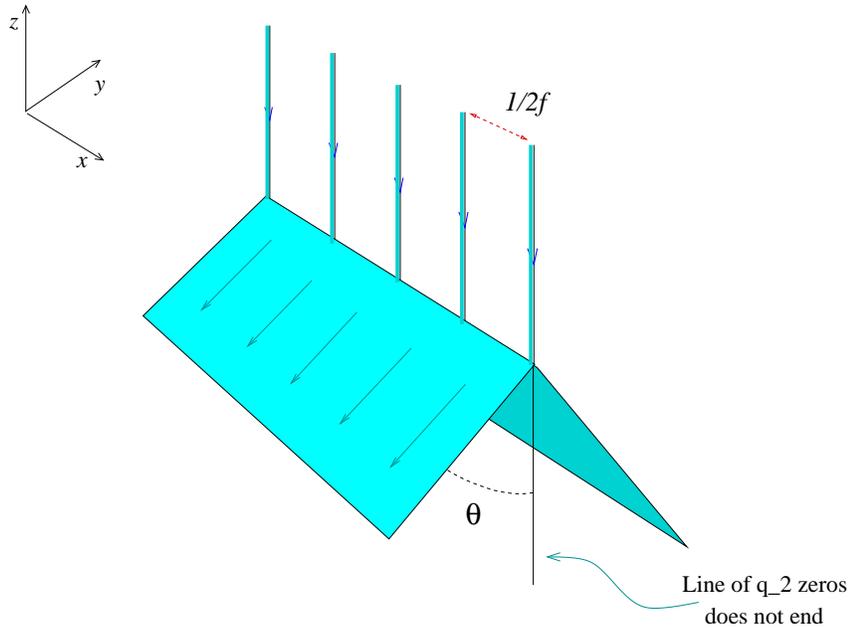}}
\caption{{\protect\small A grid of strings perpendicular to the domain wall. 
The strings form a linear lattice with distance $1/(2 f)$ between them. 
The domain wall is deformed into the shape 
shown in this figure in order to balance the tension of the strings. Half flux of 
the flux tubes goes into each of two semi-planes into which the wall is divided. 
The angle $\theta$ is determined in the text,
see Eq.~(\ref{angletheta}). It tends to $0$ when the string density $2 f$ 
is very small. It tends to $\pi/2$ when the magnetic flux 
is very large. }} 
\label{perpendicular}
\end{figure}

We can also discuss a different system that will help us to further elucidate these issues. 
Consider a grid of flux tubes perpendicular to the domain wall (Fig.~\ref{perpendicular}). The flux tubes
 are aligned along the $z$ axis and the wall, initially, is aligned in the plane $x,y$. 
We assume the flux tubes to be  equidistant  separated by intervals $1/(2 f)$. (We
consider twice the density of the previous system for later convenience). Hence, $2 f$ is the linear density. The magnetic flux density is $8 \pi f$. This simplification will allow us to make computations 
very quickly. Simultaneously, this set-up still provides us with essential information 
on physics of the ``flux tubes perpendicular to the wall'' system.   

This flux tube grid, for sufficiently large density $2 f$, 
can be considered as a surface with tension $4 \pi \xi f$ 
and linear flux $8 \pi f$. When this surface intersects 
with the domain wall we have a three-surface junction: 
the flux tube grid and  two semi-planes into which the domain wall is divided. The angle $\theta$
shown in Fig.~\ref{perpendicular}  defines geometry of the junction.  
The magnetic flux carried by the flux tubes is divided exactly into two equal parts, 
so that each semi-plane carries linear flux $4 \pi f$. 
The  wall tension  is given by 
$\xi \sqrt{ (\Delta m)^2 +  4 \pi^2 f^2 } $. 
The angle $\theta$ is determined by a simple balance of tensions,
\beq
\cos{\theta}  = \frac{2 \pi f}{ \sqrt{ (\Delta m)^2 + 4 \pi^2 f^2 }}  \, .
\label{angletheta}
\eeq
Note that $\theta = \pi/2 -\alpha$ where $\alpha$ is the angle that maximizes the BPS bound (\ref{bound}).

From this simple example we can learn an important lesson. 
First of all the flux tube grid is a source of a constant magnetic field
inside the wall. This means, as was already discussed, that the $2+1$ 
dimensional theory on the domain wall is in the Coulomb phase. Second, we can conclude that the wall with a constant magnetic field is  $1/4$ BPS
saturated. This follows from the fact that the flux tube perpendicular to the wall is known to be $1/4$ BPS (See \cite{Shifman:2002jm} and Appendix B.) 
It is straightforward to verify that, if we take the BPS equations 
for a system of strings ending on the wall (which are in Eq.~(\ref{eqboojum}))
and rotate them by an angle $\alpha$ around the $x$ axis,
we recover the same equations as those that we get from the
Bogomol'nyi completion in Eq.~(\ref{bog}).

\section{Peculiarities  of the Gauge Field Localization. Can we uplift the problem to five dimensions? }
\label{dmvqm}

As was mentioned in Sect.~\ref{CA}, 
localization of gauge fields on domain walls is not similar to that of, say, spinor field
since dualization is important. This makes the procedure 
 non-local with respect to the gauge potentials in the bulk and on the brane. 
The latter is not just a mode reduction of the former.

Let us explain it in more detail. To begin with, consider the
conventional  localization mechanism on topological defects. 
The bulk theory is defined in  space-time $X^M$ where $M=0, \dots, d-1$. The soliton is a topological object extended in $x^{\mu}$ where $\mu =0, \dots, p$. The transverse coordinates are $s^{a}$ with $a=p+1,\dots,d-1$. The theory has some bosonic fields $\phi^{(j)}(x,s)$.  The soliton is a topologically stable solution made of the bosonic fields $\phi^{(a)}$ which are independent of $x$.  The soliton is a $p$-brane spanned on the coordinates $x^{\mu}$.

Now if we want to find the spectrum of a particular scalar field 
$\varphi$ in the soliton background   we can  separate variables
\beq
\varphi(x,s)= \sum_n \varphi_{\parallel}^{(n)}(x) \varphi_{\perp}^{(n)}(s)\,.
\label{113}
\eeq  
In the quadratic in  $\varphi$ approximation the Lagrangian takes the form
\beq
\label{bulklagrangian}
L=\int d^p x \,\int d^{d-p}s  \left( \partial_M \varphi \partial^M \varphi  -  f(\phi) \varphi^2  \right)
\eeq
implying the following (linear) equations of motion
 for $\varphi$ 
\beq
(\partial_M \partial^M + f(\phi)) \varphi(X)=0\,.
\eeq 
Inserting here Eq.~(\ref{113}) we get two equations
\bea
(\partial_a \partial_a  +  f(\phi)) \varphi_{\perp}^{(n)}(s) &=& m_{(n)}^2 \varphi_{\perp}^{(n)}(s) \,,\label{transversaleq} \\[3mm]
(\partial_{\mu} \partial^{\mu} + m_{(n)}^2) \varphi_{\parallel}^{(n)}(x) &=& 0 \,,
\eea
where we used the fact that $\partial_M \partial^M = \partial_\mu \partial^{\mu} + \partial_a \partial_a$. 

The physical meaning of this formula is as follows. First we solve   the ``transverse" field equation 
 (\ref{transversaleq}) and we find the mass eigenvalues. The corresponding longitudinal field $\varphi_{\parallel}(x)$ is (generally speaking)
 a massive field in the longitudinal variables $x$. 
As long as $\varphi_{\perp}^{(n)}$ is normalizable on $s^a$ and  $m_{(n)}$ is much smaller than the inverse  soliton thickness, we can keep this soliton-localized field in the low-energy Lagrangian
emerging on the soliton world volume.
 For this particle to be massless, the corresponding transverse mode must be a zero mode of the equation (\ref{transversaleq}).

To write the effective Lagrangian for this localized field we  use the  bulk Lagrangian (\ref{bulklagrangian}), expand in longitudinal and transverse fields, and integrate 
over the the transverse variables, 
\beq
 \int ds \varphi_{\perp}(s)^2    \int d^p x  \left(  \partial_{\mu} \varphi_{\parallel}(x) \partial^{\mu} \varphi_{\parallel}(x) - m^2 \varphi_{\parallel}(x)^2   \right) .
\eeq
The norm of the perpendicular field factorizes out providing a normalization for the parallel field.
Only normalizable modes of (\ref{transversaleq}) lead to soliton-localized fluctuations.

Now let us return to 
gauge fields on 1+2-dimensional walls \cite{Shifman:2002jm}, try to
follow the way outlined above
and see that this is {\em not} the right procedure to localize
the gauge field. 

 The bulk fields we start from are the gauge field $A_{M}$, and two complex scalar fields $q_1$ and $q_2$. The bulk theory is four-dimensional,
and $q_2$. If we want to parallelize the procedure outlined above we
``nominate"
 $A_{\mu}$ ($\mu =0,1,2)$ as ``our" fields and apply the standard decomposition
\beq
\label{ordinaryway}
 A_{\mu}(X)= \varphi_{\perp}(z) A_{\mu}(x)
\eeq
(with normalizable $\varphi_{\perp}$ implying that at $z \to \pm \infty$).  Then we note that
the   magnetic field inside the wall (parallel to the wall) always involves  a derivative with over $z$. The total magnetic flux inside the wall thus contains $\int dz \partial_z \varphi_{\perp}$ and obviously vanishes.

The mode decomposition (\ref{ordinaryway}) is suitable for massive vector
fields on the wall. But there is no index theorem that can protect any zero energy solutions of equation (\ref{transversaleq}). 

What is the actual procedure leading to the gauge field localization?
We must use a global U(1) symmetry (exact in the two-flavor case or approximate in
the one flavor case). Spontaneous breaking of this symmetry localizes, through the Goldstone theorem, a phase field
on the wall. Dualization of this field gives rise to QED on the wall world volume,
with {\em electric} field directed along the wall. At the same time, the flux {\em inside}
the wall (parallel to the wall) is that of the magnetic field of the original bulk theory.

The necessity of dualization explains why
in five- (and higher dimensions) it is so hard to localize 
gauge fields on the 1+3-dimensional wall within a 
field-theoretic framework. 
 (By field theory  we mean here  something without gravity, or at least where gravity does not play essential role in the mechanism of localization.) 
 Such mechanisms could be  of enormous phenomenological interest. Ideas as to how
 one could address this problem
can be found in the literature, see e.g. \cite{Dubovsky:2001pe,two,Isozumi:2003rp}.

The construction
discussed in the present paper cannot be uplifted to higher dimensions.
Let us discuss in more detail what happens if we just lift  our four-dimensional models 
to five dimensions.
In four-dimensional theory, what is localized on the 1+2-dimensional wall is a gauge boson
\beq
\label{basicrelation}
{F}_{ij} = \epsilon_{ijk} \partial_k \sigma\,.
\eeq
where ${F}_{ij}=\partial_{[i}\widetilde{A}_{j]}$, and $\widetilde{A}_{j}$ is the localized gauge potential.

As we uplift the model to, say, five dimensions, we observe that what is localized here is not a gauge field but a $2$-form, or the Kalb--Ramond filed $\widetilde{H}_{\mu\nu}$.
Indeed, now
\beq
{C}_{\mu\nu\rho} = \epsilon_{\mu\nu\rho\sigma}\partial_{\sigma} \sigma\,,\qquad 
{C}_{\mu\nu\rho}=\partial_{[\mu} \widetilde{H}_{\nu\rho]}\,.
\eeq
To interpret this result let us ask ourselves what are the sources charged under this Kalb--Ramond field. We know that $2$-forms couple naturally to strings, or $1$-branes, much in the same way  $1$-forms couple to particles. We have, in fact, a natural candidate for this string. 
This is  the world sheet spanned   by the uplifted ANO vortices ending on the domain wall. The
ANO vortices in five dimensions are $2$-branes. This is because their codimension is fixed to be $2$ from the homotopy $\pi_1(\U(1))$. These ANO vortices ($2$-branes) end on the domain walls ($3$-branes), and the intersection between them is exactly the  string world sheet, the source of the localized Kalb--Ramond field.

\section{Conclusions}
\label{conclusion}

\begin{figure}[h!]
\epsfxsize=11cm \centerline{\epsfbox{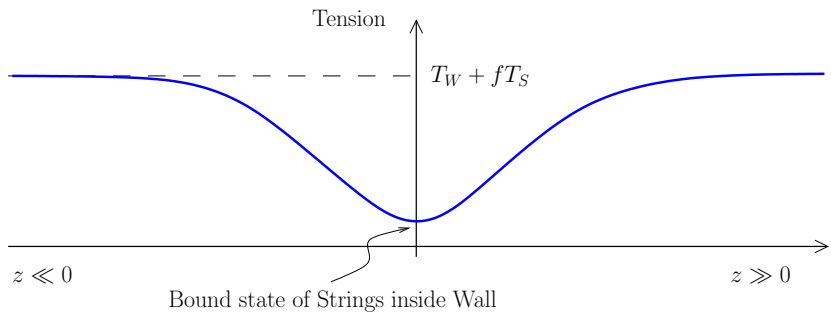}}
\caption{{\protect\small The energy of a string parallel to the wall as a function of the distance
in the one-flavor model of Sect.~2 (the minimal model).
The string on one side of the wall is the same as the string on the other side. 
The matter field condensate becomes smaller in the center of the wall;
that's why the tension of the string also reaches a minimum at this point. 
One finds a bound state of string inside the wall. }} \label{potential1flavor}
\end{figure}

Summarizing, we conclude that in certain models a string
(flux tube) inside the domain wall is possible. These are the models which we called minimal:
in particular, those in which the domain wall interpolates between two vacua where 
one and the same field condenses. The condensate is exponentially suppressed 
in the center of the domain wall. A string  in the bulk parallel to the domain wall 
is attracted to the wall. The lowest-energy configuration 
is achieved when the zeroes of the field are at $z=0$. Figure~\ref{potential1flavor}
illustrates this example. 
As a result, the U(1) filed trapped on the wall is in the confinement regime.
 
\begin{figure}[h!]
\epsfxsize=11cm \centerline{\epsfbox{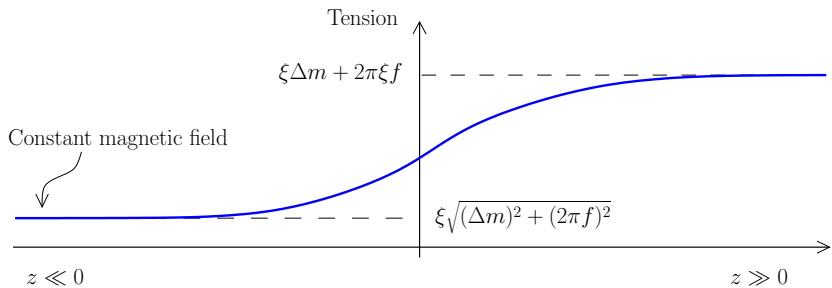}}
\caption{{\protect\small 
In the two-flavor model discussed in Sect.~4,
we consider a domain wall and a set of parallel flux tubes with density 
$f$  placed at a certain distance $z$ form the wall. 
At large positive $z$ the tension of this configuration is given by the sum of the tensions
of the wall and the grid of the flux tubes. 
The energy minimum is reached when the distance (defined as the position of the zeroes) 
is $-\infty$. Then the configuration looks as the domain wall with a constant magnetic field on it. }} 
\label{potential2flavors}
\end{figure}

We compared this regime with that of the two-flavor model.
The 
lowest-energy configuration here is achieved
when the  zeroes are at $z=-\infty$ (the mirror reflected
solution $z\leftrightarrow -z$ is also possible). Now, the Coulomb phase on the wall is energetically preferred to the confining phase (see Fig.~\ref{potential2flavors}).
Underlying dynamics is not so transparent as in the minimal model, with the ``confinement on the wall" regime. Technically the distinction is due to the fact that in the two-flavor model  $q_1$ vanishes on one side of the wall and $q_2$ on the other,
so there is no relative phase between, say, $  q_1$ on the left of the wall and $q_1$ on the right
of the wall.
There is only a phase between $  q_1$ on the left and $q_2$ on the right, and it is not $z$ dependent.

This technical argument is backed up by an iron-clad symmetry argument.
The two-flavor model has a global U(1) in the bulk, which is spontaneously
broken on the wall. The Goldstone boson
of this breaking is the $\sigma$ field. It is strictly massless. On the other hand,
one can trace its connection to the photon of the bulk theory which is Higgsed in the 
bulk \cite{Shifman:2002jm}.

The mechanisms we discussed in this paper are quite general. We presented two working examples: a toy minimal model in Sect.~\ref{toymodel} and a strong coupling example in Sect.~\ref{seibergwitten}.  In both examples there is a condensate which does not vanish in
two distinct vacua separated by the domain wall, and a residual (suppressed)
condensate  in the middle of the wall. The quanta of the  fields which condense in both vacua
are very heavy inside the wall; hence, one can view the residual condensate
as a ``tunneling effect" \cite{Dvali:1996xe}.

\section*{Acknowledgments}

We want to thank G. Dvali for correspondence and useful discussions.
R.A. is grateful to FTPI for their hospitality in February 2008 when a part of this work was done.
S.B. wants to thank Ki-Myeong Lee and people at KIAS for hospitality
extended to him in June 2008 
and for interesting discussions. This paper was presented 
by S.B. at the conference CAQCD-08 in Minneapolis in May 2008.

This work is supported by DOE grant DE-FG02-94ER40823.
The work of M.S. was supported in part by  {\em Chaire Internationalle de Recherche Blaise
Pascal} de l'Etat et de la R\'{e}goin d'Ile-de-France, g\'{e}r\'{e}e par la 
Fondation de l'Ecole Normale Sup\'{e}rieure.
The work of A.Y. was  supported 
by  FTPI, University of Minnesota, 
by RFBR Grant No. 09-02-00475a,
and by Russian State Grant for 
Scientific Schools RSGSS-11242003.2.
\newpage

\section*{Appendix A. Vacuum structure}   
\label{vacua}
  \addcontentsline{toc}{section}{Appendix A. Vacuum structure}

\renewcommand{\theequation}{A.\arabic{equation}}
\setcounter{equation}{0}

This appendix is devoted to exact computation of the vacuum structure of the theory
discussed  in 
Sect.~\ref{seibergwitten} using the technique developed in \cite{Cachazo:2002zk}. The
 U(2)  $\mathcal{N}=2$ gauge theory,
without hypermultiplets, has the following Seiberg--Witten curve
\bea
\label{vv}
{y}^2&=& (z-\phi_1)^2(z-\phi_2)^2-  \Lambda^{4} \nonumber\\[3mm]
&=& (z^2 -u_1 z + \frac{u_1^2}{2}-u_2)^2-  \Lambda^{4} \,.
\eea
The moduli space consists of the Coulomb branch parametrized by $u_1 =\Tr \phi$ and $u_2 = \frac12 \Tr \phi^2$. The four roots of the curve are
\bea
z^-_{1,2} &=&  \frac{u_1}{2} \pm \sqrt{ -\frac{u_1^2}{4} + u_2 +  \Lambda^2}  \ ,  \nonumber \\[3mm]
 z^+_{1,2}&=& \frac{u_1}{2} \pm \sqrt{ -\frac{u_1^2}{4} + u_2 -   \Lambda^2)}  \, .
\eea
Since we have no hypermultiplets the singularity structure is very simple and consists just of the monopole and dyon singularities (co-dimension $2$ surfaces) without intersections
\beq
\label{singularities}
 u_2= \frac{u_1^2}{4} \pm  \Lambda^2 = 0
\eeq

Now, let us   break extended supersymmetry down to $\N=1$ 
by virtue of the superpotential 
\beq 
W= \alpha \, \Tr \, \left(
\frac{\Phi^3}{3} - \xi \Phi \right). 
\eeq 
Classically, we have the three vacua (\ref{threevacua}).  
The quantum solution goes as follows. The Seiberg--Witten curve factorizes,   
\beq
y^2 = P_N^2 -  \Lambda^{2N} = F_{2n} H_{N-n}^2\,,
\eeq   
where $n$ is the number of the unbroken U(1) factors in the low-energy theory. This factorization is then related to the superpotential parameters,
\beq
\label{polynomial}
y_m^2 = W_k^{\prime 2} + f_{k-1} = F_{2n} \widetilde{Q}_{k-n}^2
\eeq
with $$\widetilde{Q}_{k-n} = V_{k-N} H_{N-n} + Q_{N-n-1}\,.$$ 
The unknown parameters are the coefficients of $f_{k-1}$, 
the coefficients of 
$V_{k-N}$ and $Q_{N-n-1}$, and, finally
$n$ parameters of the U$(1)^n$ Coulomb moduli space. In total $k+n+(k-N)+(N-n-1) = 2k -1$, exactly the number of equations from the polynomial equality (\ref{polynomial}).

In the case of interest $N=2$,  $k=2$  and $n=1,2$ depending on whether we deal with the confining vacua or the Coulomb one. The solutions are as follows.

\vspace{2mm}

 {\it (i) Monopole}

\noindent
 We must lie in the monopole singularity
$$u_2= \frac{u_1^2}{4} +  \Lambda^2 = 0\,.$$
On this surface the factorization equation is 
\beq
y_m^2 = (z^2-\xi)^2 + f_{1} =
 \left(z^2 + -u_1 z +\frac{u_1^2}{4} - \Lambda^2\right) 
 \left(z-\frac{u_1}{2} - a \right) ^2\,.
\eeq
The solution is 
\beq
f_{1} = \mp 4 \Lambda^2 \sqrt{\xi - \Lambda^2 } z + 4\Lambda^2 \xi + 3 \Lambda^4 \,\,\,{\rm and}
\,\,\,
 a= u_1 = \pm    2\sqrt{\xi - \Lambda^2 }\,,
\eeq 
where $\pm$ correspond to  two classical non-Abelian vacua in Eq.~(\ref{classicalvacua}).

\vspace{2mm}

 { \it (ii) Coulomb} 

\noindent 
  The factorization in this case is  very simple 
\beq
y_m^2 = (z^2-\xi)^2 + f_{1} = P_2^2 -  \Lambda^{4}\,.
\eeq
The solution is
$u_1=0$, $u_2=\xi$ and  $f_1 = - \Lambda^4$. There is no change from the classical formula.

\vspace{2mm}

 {\it (iii) Dyon}
 
\noindent 
We proceed in the same as above in the monopole case, but now we lie in the singularity
$$u_2= \frac{u_1^2}{4} -  \Lambda^2 = 0\,.$$ The factorization is 
\beq
y_m^2 = (z^2-\xi)^2 + f_{1} = (z^2 + -u_1 z +\frac{u_1^2}{4}  + 2 \Lambda^2) (z-\frac{u_1}{2} - a ) ^2\,.
\eeq
The solution is 
\beq
f_{1} =  \pm 4 \Lambda^2 \sqrt{\xi + \Lambda^2 } z - 4\Lambda^2 \xi + 3 \Lambda^4 
\,\,\, {\rm   and}
\,\,\, 
 a= u_1 = \pm 2\sqrt{\xi + \Lambda^2 }\,.
\eeq

\vspace{2mm}

Summarizing, we have  five vacua, in total. One is the Coulomb vacuum, 
whose position in the moduli space is not modified by quantum corrections. 
Four others are two monopole  and two dyon vacua. Their $u_2$ coordinate 
is not modified by quantum correction, but the $u_1$ coordinate is changed. 
This is why they are aligned in Fig.~\ref{vacuaxibig}. At the critical value 
$\xi  =  \Lambda ^2$ the two monopole vacua and the Coulomb vacuum 
coalesce together. In Sect.~\ref{seibergwitten} we performed
our analysis
near this critical value, in order to have a low-energy effective 
action which is weakly coupled on the domain wall profile. Another 
critical value is at $\xi =-\Lambda^2$ were the dyon vacua 
coalesce with the Coulomb one.

\section*{Appendix B. Comments on supercharges}
\label{supercharges}
  \addcontentsline{toc}{section}{Appendix B. Comments on supercharges}

\renewcommand{\theequation}{B.\arabic{equation}}
\setcounter{equation}{0}

In this appendix we will show, basing on  the central charges,
that in $\mathcal{N}=2$ SQED   configurations
with flux tubes parallel to the domain wall are not BPS saturated.
We will follow the formalism of Ref.~\cite{Shifman:2002jm}.

The supersymmetry transformations in $\mathcal{N}=2$ SQED are
given by the following expressions (where  $f,p=1,2$ are SU$(2)_R$ indices):
\begin{eqnarray}
&&
\delta\lambda^{f\alpha}=\frac12(\sigma_\mu\bar{\sigma}_\nu\ve^f)^\alpha
F_{\mu\nu}+\ve^{\alpha p}D^a(\tau^a)^f_p
+i\sqrt{2}\partial\hspace{-0.65em}/^{\,\,\alpha\dot{\alpha}}\, a \, 
\bar{\ve}^{f}_{\dot\alpha}\ ,
 \nonumber\\[2mm]
&& \delta\psi^{\alpha A}\ =\ i\sqrt2\
\nabla\hspace{-0.65em}/^{\,\,\alpha\dot{\alpha}}q^{fA}\bar{\ve}_{f\dot{\alpha}}
+\sqrt{2}\ve^{\alpha f}F_f^{A}\ ,
\nonumber\\[2mm]
&& \delta\tilde{\psi}^{\alpha}_{A}\ =\ i\sqrt2\
\nabla\hspace{-0.65em}/^{\,\,\alpha\dot{\alpha}}\bar{q}^f_{A}\bar{\ve}_{f\dot{\alpha}}
+\sqrt{2}\ve^{\alpha f}\bar{F}_{fA}\ ,
\label{str}
\end{eqnarray}
 $D^a$ is the SU(2)$_R$ triplet of $D$ terms:
\beq
\label{dterm}
D^1=i\frac{g^2}{2}\left(|\vp^A|^2-\xi\right),\; D^2=D^3=0\,,
\eeq
while $F^f$ and $\bar{F}_f$ are the matter $F$ terms,
\beq
\label{fterm}
F^{fA}=i\frac1{\sqrt{2}}\left(a+\sqrt{2}m_A\right)q^{fA},\qquad
\bar{F}_{Af}=i\frac1{\sqrt{2}}\left(\bar{a}+\sqrt{2}m_A\right)\bar{q}_{Af}.
\eeq

Let us consider a BPS domain wall, oriented in the  $(x,y)$ plane.
The following supersymmetry transformations are left unbroken by the domain wall:
\begin{eqnarray}
\bar{\ve}^2_{\dot{2}}&=&-i\ve^{21},\quad
\bar{\ve}^1_{\dot{2}}=-i\ve^{22},
\nonumber\\[3mm]
\bar{\ve}^1_{\dot{1}}&=&i\ve^{12},\qquad
\bar{\ve}^2_{\dot{1}}=i\ve^{11}.
\label{susy1}
\end{eqnarray}

Let us consider a vortex parallel to the $z$ axis.
The following supertransformations are left unbroken:
\begin{eqnarray}
&& \ve^{12} =- \ve^{11}\, , \quad \bar{\ve}_{\dot{1}}^{2} =- \bar{\ve}_{\dot{1}}^{1}   \,,  
    \nonumber\\[3mm]
&& \ve^{21} =\ve^{22}\, , \qquad \bar{\ve}_{\dot{2}}^{1} =\bar{\ve}_{\dot{2}}^{2}\, .
\label{susy2}
\end{eqnarray}

It is possible then to find a $1/4$-BPS soliton corresponding
to a vortex perpendicular to the wall, because Eqs.~(\ref{susy1},\ref{susy2})
are compatible in the sense that we can solve both  constraints taking
\begin{eqnarray}
\label{epsilon}
&& \ve^{12} =- \ve^{11}\ ,  \nonumber\\[2mm]
&& \ve^{21} =\ve^{22},
\end{eqnarray}
and the $\bar{\epsilon}$ given by Eq.~(\ref{susy1}). 

Let us then consider a vortex parallel to the $y$ axis.
From a simple spinor rotation, we can find
supertransformations left unbroken,
\begin{eqnarray}
&& \ve^{12} =- i \ve^{21}\, , \quad
 \bar{\ve}_{\dot{1}}^{2} =- i \bar{\ve}_{\dot{2}}^{1}   \,,  
    \nonumber\\[3mm]
&& \ve^{11} =-i \ve^{22}\, , \qquad \bar{\ve}_{\dot{1}}^{1} =-i \bar{\ve}_{\dot{2}}^{2}\, .
\label{susy3}
\end{eqnarray}
It is easy to check that it is not possible to find a non-trivial 
solution to  the constraints in Eqs.~(\ref{susy1}) and (\ref{susy3}).
This shows that the configuration with the flux tube parallel to the domain wall
breaks all  supersymmetries of the theory.

Let us then consider a domain wall in the $(x,y)$ plane with some constant
magnetic field along the $x$ axis.
As discussed in Figure~\ref{perpendicular}, the unbroken supercharges are the same of
the system of a vortex ending on a wall (modulo a rotation
  by the angle in Eq.~(\ref{angletheta}) along 
the $y$ axis).


\end{document}